\documentclass[11pt]{article}  
\usepackage{graphicx}
\usepackage[a4paper,left=3cm,right=3.cm, bottom=2.cm, top=2.0cm]{geometry}
\usepackage{amssymb}
\usepackage{amsmath}
\usepackage{bm}
\usepackage[margin=1pt,font=small,labelfont=bf]{caption}
\usepackage[dvipsnames]{xcolor}
\definecolor{citecolor}{RGB}{128,0,32}
\definecolor{headcolor}{RGB}{128,128,128}
\usepackage[unicode,hyperfootnotes=false,breaklinks=true,colorlinks=true,allcolors=citecolor]{hyperref}
\usepackage{setspace}
\usepackage{fancyhdr}
\usepackage{titlesec}
\usepackage{algorithm}
\usepackage{algpseudocode}
\usepackage{etoolbox}

\usepackage{libertinus}
\usepackage{tabularx}
\usepackage{datetime} 
\usepackage{lineno}
\usepackage{etoolbox}
\usepackage{subcaption} 
\patchcmd{\linenumber}{\hb@xt@}{\hbox}{}{}
\usepackage[ giveninits=true, style=authoryear, backend=biber, maxcitenames=2, maxbibnames=99,
  hyperref=true, backref=false, sorting=nyt, uniquename=init, autolang=hyphen,
  dashed=false]{biblatex}

\AtEveryCite{\it \color{citecolor}} \AtEveryBibitem{\clearfield{month}}
\AtEveryBibitem{\clearfield{issn}}
\renewbibmacro{in:}{}

\let\citeA\textcite
\let\cite\parencite
\DeclareGraphicsExtensions{.jpg} \graphicspath{{./figs/}}
\newdateformat{usvardate}{\shortmonthname[\THEMONTH]. \THEDAY, \THEYEAR}
\newcommand{\lastmodified}{%
  \textcolor{headcolor}{\scriptsize \usvardate\today{} at \currenttime} }
\fancypagestyle{custom}
{      
\fancyhead[L]{}
\fancyhead[C]{\textcolor{headcolor}{\scriptsize submitted to \textit{AGU Advances}}}
\fancyhead[R]{}
\fancyfoot[C]{\textcolor{gray}{\thepage}}
\fancyfoot[R]{\lastmodified}
\fancyfoot[L]{}
}
 \titlespacing\section{0pt}{12pt plus 4pt minus 2pt}{0pt
plus 2pt minus 2pt}

\begin{document}

\pagestyle{custom}
\begin{center}
\LARGE {\bf Coupled Transient Processes Govern Intraplate Earthquake Swarm Evolution: Insights from the 2019-2020 Palghar Sequence\\[12pt]}

\normalsize
Ratna Bhagat$^{1,2}$,
Pathikrit Bhattacharya$^{1,2,\P}$,
K. M. Sreejith$^{3}$,
Harsha S. Bhat$^{4}$,
Vineet K. Gahalaut$^{5,6}$
\\[12pt]

\begin{enumerate}
	\scriptsize
	\setlength\itemsep{-5pt}
	\item {National Institute of Science Education and Research, Jatni, Odisha 752050, India}
    \item {Homi Bhabha National Institute, Training School Complex, Anushakti Nagar, Mumbai 400094, India}
    \item {Geosciences Division, Space Applications Centre, Indian Space Research Organisation, Ahmedabad 380015, India}
    \item {Laboratoire de Géologie, École Normale Supérieure, CNRS UMR 8538, PSL Research University, 75005 Paris, France}    
    \item {Wadia Institute of Himalayan Geology, Dehradun, Uttarakhand 248001, India}    
    \item {CSIR--National Geophysical Research Institute, Hyderabad 500007, India}    
\end{enumerate}

\let\thefootnote\relax\footnotetext{$\P$ Corresponding author: \texttt{pathikritb@niser.ac.in}}
\end{center}

\section*{Key Points}
\begin{itemize}
  \small
  \item A machine-learning-enhanced earthquake catalog resolves two normal faults within granitic basement and associated off-fault seismicity.
  \item Apparently diffusive swarm envelope emerges from the superposition of intermittent slow and fast migration episodes.
  \item Sequential fault activation and coordinated seismicity migration consistent with fluid migration and aseismic-slip mediated stress transfer.
\end{itemize}

\section*{Abstract}
\small

Earthquake swarms provide a natural window into fault response to transient perturbations, yet their driving processes are commonly interpreted using fluid-driven and aseismic-slip-driven end-member models. Intraplate swarms offer a unique setting to test these models because low secular tectonic loading heightens the sensitivity of faults to transient, non-tectonic forcing. We use a machine-learning-enhanced catalog of $\sim$50,000 earthquakes from the 2019–2020 Palghar earthquake swarm in western India to test this end-member framework. High-resolution relocations, together with moment tensor solutions, reveal two shallow normal faults in a granitic basement, with seismicity sequentially migrating from the western to the eastern fault before expanding into the intervening damage zone. Although the swarm exhibits an overall diffusion-like expansion, the relocated seismicity reveals a persistent $\sim$5-km-deep localized seismicity band and repeated migration fronts sometimes propagating faster than expected from fluid diffusion alone. The velocity–duration scalings of these intermittent episodes span both fluid- and slow-slip-driven regimes. Sequential fault activation and contrasting migration styles throughout the swarm duration reveal dynamics that cannot be explained by either end-member mechanism alone. Instead, the Palghar swarm evolved through coupled fluid-assisted deformation and transient stress transfer within an interacting fault network, with migration episodes consistent with aseismic deformation. These observations reveal that high-resolution catalogs can disentangle transient processes hidden within apparently diffusive swarm behavior. More broadly, intraplate earthquake swarms provide powerful natural laboratories for resolving how coupled transient processes govern earthquake triggering and fault interaction in stable continental crust.

\section*{Plain Language Summary}

Earthquake swarms represent clusters of seismicity that occur over days to years without a single large mainshock. Swarms within plate interiors, where the crust is only weakly influenced by tectonic forces, are especially useful for understanding what non-tectonic forces make faults slip. Recent studies have proposed that swarms generally fall into two categories: those driven by subsurface fluids, and those driven by slow fault slip. We tested this idea using the 2019-2020 Palghar earthquake swarm in western India, analyzing recordings from two local seismic networks with a machine-learning method that identified nearly 50,000 earthquakes. We found the swarm activated two shallow, preexisting faults and the fractured rock between them, extending to about 8 km depth. Activity began on one fault, spread to the second, and then expanded between them as the faults increasingly interacted - at times migrating faster than fluid movement alone could explain. These findings suggest the swarm was driven not by fluids or aseismic slip alone, but by a combination of both plus stress transferred between the faults, challenging the view that swarms fit neatly into one category. This shows faults in low-stress, intraplate settings can be highly sensitive to such combined, temporary triggers.

\section{Introduction}\label{intro}

Earthquake swarms are clusters of seismicity that occur close together in space and time without a clearly identifiable initiating event \cite{mogi1962magnitude, hill1977model, vidale2006survey}. Swarm activity commonly exhibits complex spatiotemporal migration of seismicity, reflecting the combined influence of multiple physical processes rather than a single triggering mechanism \cite{hainzl2004seismicity, roland2009earthquake, ruhl2016complex, danre2022prevalence, fischer2023fast, danre2024parallel}. Characterizing these migration patterns, therefore, provides an important means of distinguishing among the processes responsible for swarm evolution. For example, slow, diffusive migration is commonly interpreted as reflecting fluid-mediated processes wherein the seismicity front is phenomenologically linked to the migration of a pore-pressure diffusion front \cite{shapiro2009fluid, fischer2021growth, danre2022prevalence}. In contrast, faster-than-diffusive migration may reflect elastic stress transfer arising from a wide range of processes, including poroelastic responses to fluid flow within deformable porous media, the rapid propagation of aseismic slip fronts, and stress redistribution through elastic interactions within actively deforming fault networks \cite{hill1977model, rao1991earthquake, hainzl2004seismicity, goebel20172016, bhattacharya2019fluid, hatch2020evidence, danre2024parallel, almakari2026fault}. Determining the processes that govern swarm evolution is important for understanding earthquake triggering, the physical controls on the spatiotemporal evolution of earthquake clusters, and the partitioning of strain-energy release across event sizes \cite{vidale2006survey,  roland2009earthquake, hatch2020evidence, ross20203d,  danre2024parallel}. Intraplate swarms are particularly valuable in this context because they provide a rare window into fault responses to transient stress perturbations under low long-term strain rates, where transient perturbations to fault stability can exceed the secular strain accumulation in amplitude \cite{hill1977model, ellsworth2013injection}.

Many intraplate swarms are triggered by transient stress or strength perturbations and are often associated with pre-existing zones of crustal weakness \cite{rastogi1997seismicity, stevenson2006booming, srinagesh2020appraisal}. Within the stable continental crust of India, several such swarms have been reported, but most are short-lived and poorly monitored, limiting insights into their driving mechanisms \cite{srivastava1996comparison, sateesh2019earthquake, wadhawan2021monsoonal}. Of these, the Palghar swarm is an exception in that it continued for over two years and produced, as we show later, many tens of thousands of events and was well monitored by multiple seismic networks deployed by government agencies \cite{srinagesh2020appraisal, gahalaut2022long, sharma2023characteristics}. Most of the seismic activity occurred within the granitoid basement of the Deccan Volcanic Province within a relatively simple fault system that enables clear interpretation of fault–seismicity relations \cite{sharma2020long, pavankumar2020magnetotelluric, nath2021dynamic}. Previous studies, based on smaller, local seismic networks and multiple geophysical observations, have suggested the presence of fluids within the fractured upper crust in the Palghar swarm region \cite{pavankumar2020magnetotelluric, nath2021dynamic, subhadra2024attenuation, kanaujia2025indication}. However, substantial discrepancies in hypocentral locations between the seismicity catalogs derived from two major local networks limit our ability to resolve the swarm's migration patterns and, consequently, to identify the physical processes controlling its initiation and evolution \cite{sharma2020long, srinagesh2020appraisal}. As a result, the relative roles of pore-pressure diffusion, aseismic slip (if any), and fault-seismicity interactions, as well as the possible source of the fluids, all remain poorly constrained.

Resolving the contribution of these processes requires a dense catalog with accurate earthquake locations, as precise hypocenters and source-property estimates are essential for diagnosing swarm driving mechanisms. To achieve this, we merge waveform data from two local seismic networks to construct a region-specific velocity model and a unified, high-resolution earthquake catalog for the Palghar swarm. We use a machine-learning-based detection workflow to substantially densify the earthquake catalog and refine earthquake locations using relocations based on waveform cross-correlation derived differential travel times. The resulting catalog resolves the three-dimensional geometry and spatiotemporal evolution of the Palghar swarm, revealing two closely spaced fault zones and deformation within an intervening damage zone. We show that seismicity migrated between the two faults through a combination of diffusion-like expansion and transient episodes of rapid propagation, indicating the combined influence of elastic fault interactions, crustal fluids, and transient aseismic slip. These observations demonstrate how fault interactions and transient deformation govern the evolution of intraplate earthquake swarms and provide new insight into earthquake triggering under low background strain-rate regimes.

\section{Geological setting, data and methods}

The Palghar region lies on the western margin of the Deccan Volcanic Province (DVP), where $\sim$1.5 km thick basalt flows overlie Archaean–Proterozoic basement rocks and are dissected by inherited rift-related structures, including the Panvel flexure, the West Coast Fault, and several N–S trending basement-rooted lineaments \cite{auden1949dykes, courtillot1986deccan, courtillot2003three, chandrasekharam1985structure, gunnell1998shoulder,   sheth1998reappraisal, nath2021dynamic, sharma2020long, sharma2023characteristics}. These inherited structures have hosted diverse modes of seismicity in the past, from the 1993 Latur earthquake and reservoir-triggered seismicity at Koyna to recurrent earthquake swarms \cite{rao1991earthquake, srivastava1996comparison, raju2000micro, hainzl2014monsoon, sateesh2019earthquake, wadhawan2021monsoonal}. 

Among these swarms, the Palghar sequence stands out for its unusually long duration (over two years), high event count (tens of thousands of earthquakes), and dense instrumental coverage, providing a rare opportunity to constrain—if not the triggering mechanisms—at least the processes governing its evolution \cite{sharma2020long, sharma2023characteristics, gahalaut2022long}. Previous studies have linked the Palghar swarm to both meteoric-water infiltration and deeper crustal fluids \cite{pavankumar2020magnetotelluric, srinagesh2020appraisal, sharma2020long, nath2021dynamic, gahalaut2022long,  kanaujia2025indication}. Its exceptional duration and productivity provide a unique opportunity to investigate the processes governing earthquake clustering and fault reactivation within the Deccan Volcanic Province, with broader implications for intraplate seismicity worldwide.


\subsection{\label{sec:data}Seismic Data}

This study uses waveform data from 11 broadband stations operated independently by the National Geophysical Research Institute (NGRI) and the National Center for Seismology (NCS) following the onset of the Palghar swarm, providing dense coverage over an area of approximately 30 km × 30 km \cite{sharma2020long, gahalaut2022long} (SI Figure~\ref{fig:intro_map}). We merge the independently compiled NGRI and NCS catalogs by matching events within 0.5 s of their origin times, yielding a unified dataset with an improved mean azimuthal gap ($\sim$80°) compared with the individual NCS ($\sim$158°) and NGRI ($\sim$94°) networks (SI Figure~\ref{fig:gap_comparison}). Earthquakes recorded by at least ten stations are used to derive a one-dimensional velocity model (Section~\ref{sec:velocity_model}), which is subsequently used for earthquake detection, absolute locations, relative relocations, and moment tensor inversions. All remaining catalog construction steps rely on automated detection and relative location procedures. Station coordinates and elevations are listed in SI Table~\ref{tab:station_info}.

\subsection{\label{sec:velocity_model}One-Dimensional Velocity Modeling}

Previous studies adopted different velocity models for the NGRI and NCS datasets, resulting in substantial discrepancies in earthquake depths (6-7 km in \citeA{sharma2020long} to 4-16 km in \citeA{srinagesh2020appraisal}). To obtain a consistent velocity model for the merged dataset, we applied joint hypocenter determination (JHD) using \texttt{VELEST} \cite{kissling1995program}. The inversion used 2,514 well-recorded human-picked earthquakes observed at ten or more stations, yielding 25,097 P- and 24,361 S-wave arrivals. A constant $V_p/V_s$ ratio of 1.71, estimated from the linear relationship between P- and S-wave arrival times (SI Figure~\ref{fig:vp_vs}), was assumed throughout the inversions.

\begin{figure}[ht]
    \centering
    \includegraphics[width=1\linewidth]{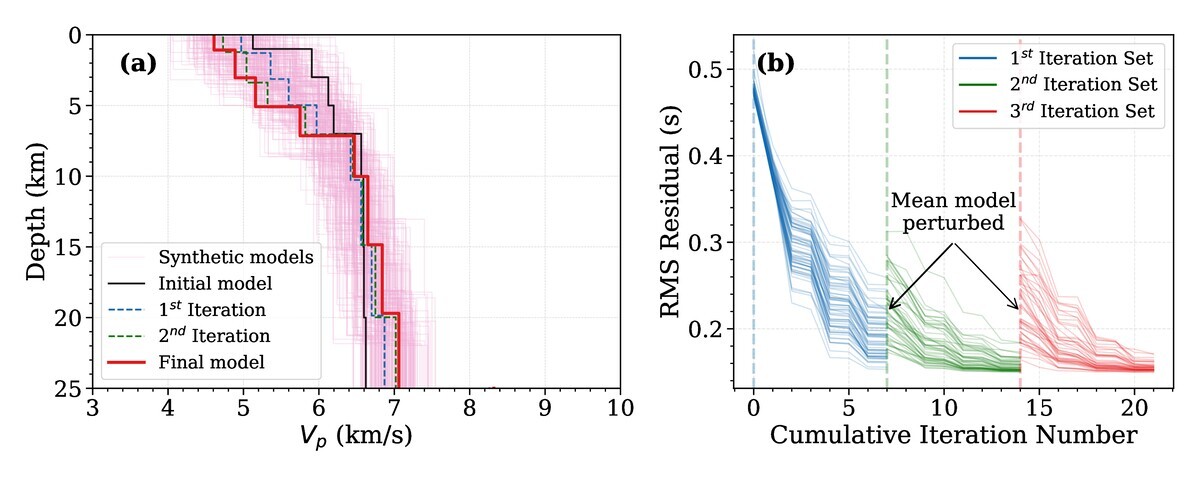}\caption{Joint hypocenter–velocity inversion results for the Palghar swarm region using \texttt{VELEST}. (a) Velocity inversion results after three inversion stages. The initial velocity model of \citeA{srinagesh2020appraisal} is shown in black. At each stage, an ensemble of velocity models was generated by randomly perturbing the final mean model from the previous stage by $\pm0.7$ km s$^{-1}$ in layer velocity and $\pm2$ km in interface depth to assess inversion stability. The resulting ensemble of final velocity models across all 3 stages is shown in pink, and the mean at the end of the stage 3 (thick red line) was adopted in this study. Blue and green dashed lines denote the mean velocity models obtained after the first and second inversion stages, respectively. (b)  Evolution of RMS travel-time residuals during the inversion. Vertical dashed lines indicate the boundaries between the three inversion stages.}
    \label{fig:velocity_model}
\end{figure}

The one-dimensional model of \citeA{srinagesh2020appraisal} was used as the initial reference. To assess model stability and minimize dependence on the starting model, we generated ensembles of randomly perturbed velocity models by varying layer velocities ($\pm0.7$ km s$^{-1}$) and interface depths ($\pm2$ km) within prescribed ranges. Three successive inversion stages (see SI materials for details) comprising 210 perturbed models in total converged to an ensemble of solutions at the end of the third stage (Figure~\ref{fig:velocity_model}). The mean velocity model from this final ensemble was adopted for all subsequent earthquake detection, location, and moment tensor analyses. Station corrections estimated during the inversion account for near-surface heterogeneity and were applied in subsequent relocations.

\subsection{ML-based earthquake detection and relocation}
Further earthquake detection and hypocentral determination were performed using a multi-stage, fully automated workflow designed to construct a dense and internally consistent earthquake catalog for the Palghar swarm. The workflow combines network-based detection using the BPMF ({\textbf B}ack {\textbf P}rojection and {\textbf M}atched {\textbf F}iltering) framework \cite{beauce2024bpmf}, probabilistic absolute location, and high-precision relative relocation, without reliance on manual phase picking for catalog construction.

\begin{figure}[ht]
    \centering
    \includegraphics[width=\linewidth]{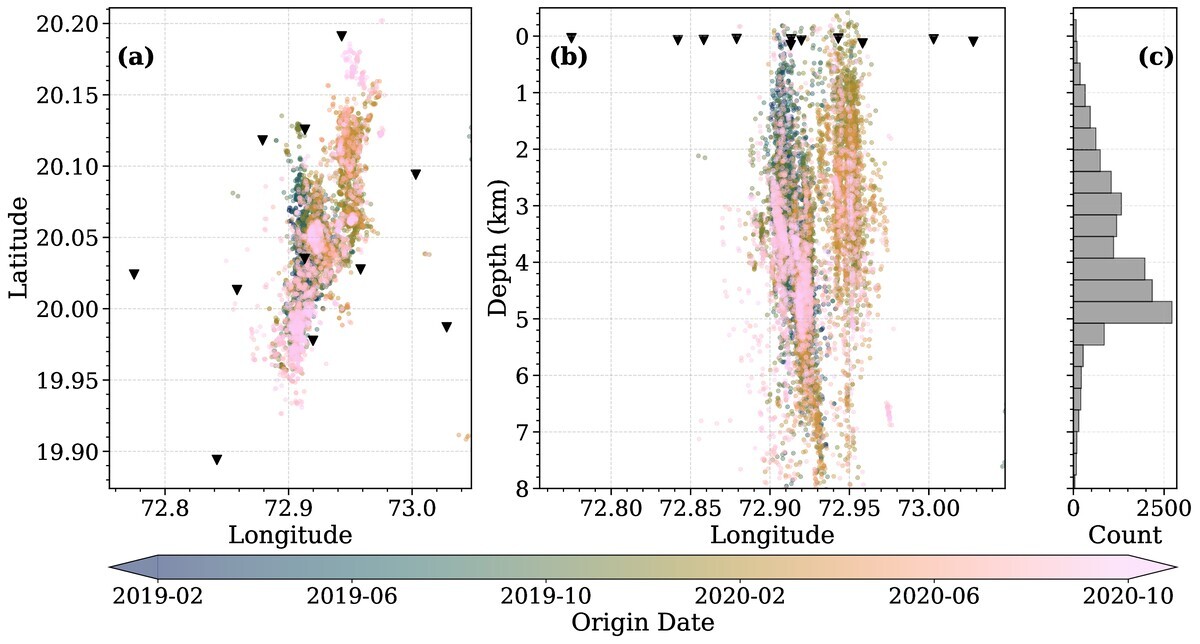}
    \caption{
    Spatial distribution of the \texttt{GrowClust}-relocated Palghar earthquake swarm.
    (a) Map view, and
    (b) longitude–depth cross-section of relocated seismicity color-coded by origin time, illustrating the spatiotemporal evolution of the swarm. Black inverted triangles denote seismic station locations projected onto the section. (c) Histogram of hypocenter depths showing that seismicity is predominantly concentrated between about 2 and 6 km. The flood basalt is expected to be not more than 1.5 km thick in this region \cite{sharma2023characteristics}. Therefore, the seismicity is almost entirely concentrated within the granitoid basement.
    }
    \label{fig:depth_section}
\end{figure}

Continuous three-component waveform data were first processed using the deep-learning phase picker PhaseNet \cite{zhu2019phasenet} to estimate P- and S-wave arrival probabilities. These probabilities were used in the back projection workflow developed by \cite{beauce2024bpmf}, which coherently stacked phase-consistent arrival probabilities using theoretical moveout corrections computed from the one-dimensional velocity model (Section~\ref{sec:velocity_model}). Events exceeding a predefined beam-power threshold (SI Figure~\ref{fig:beam_power}) yielded an initial catalog of 57,826 earthquakes.

These initial grid-based hypocenters obtained from the back projection procedure were subsequently refined using the probabilistic location algorithm \texttt{NonLinLoc} \cite{lomax2000probabilistic}. Retaining only events recorded at six or more stations produced 20,638 earthquakes with median horizontal and depth uncertainties of 0.5 km and 1.5 km, respectively (SI Figure~\ref{fig:NC_GC}). High-precision relative locations were then obtained using \texttt{GrowClust} \cite{trugman2017growclust} and waveform cross-correlation of 5–10 Hz filtered seismograms. Event pairs sharing at least six observations with cross-correlation coefficients $\geq0.8$ were retained, resulting in a final catalog of 16,008 earthquakes. For comparison, applying the same \texttt{GrowClust}  relocation workflow to the manually picked catalog results in only 5,765 relocated events.

The relocated seismicity delineates two narrow, subparallel north–south–striking fault strands confined to depths of ~2–6 km (Figure~\ref{fig:depth_section}). Bootstrap analysis indicates median relative location uncertainties of ~250 m horizontally and ~320 m vertically (SI Figure~\ref{fig:NC_GC}), sufficient to robustly resolve fault geometry and swarm evolution.

\begin{figure}[h!]
    \centering
    \includegraphics[width=\linewidth]{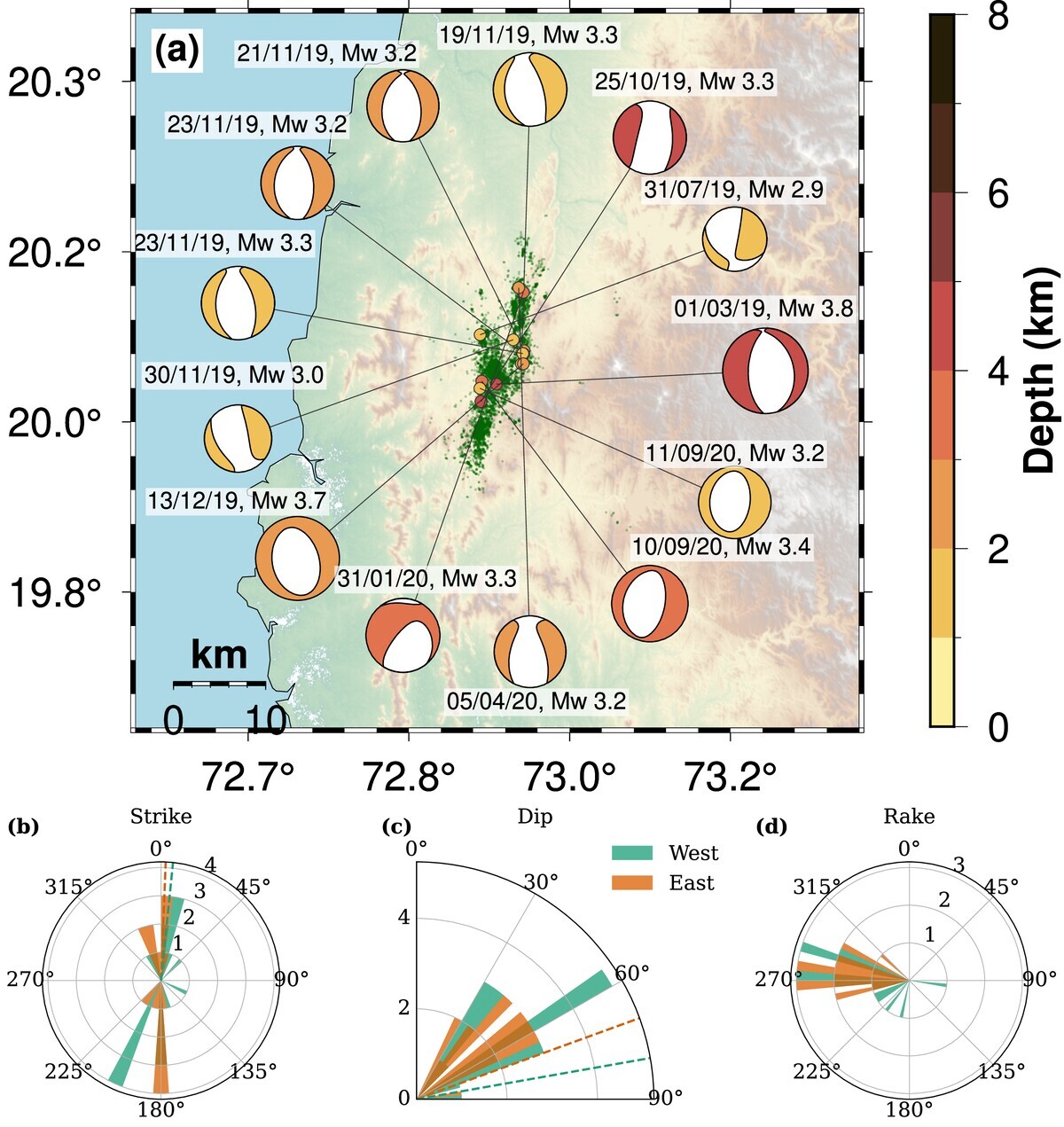}
    \caption{Moment tensor solutions and fault-plane orientation statistics for 37 earthquakes (\mbox{$M_l \geq 3.0$}) in the Palghar swarm. (a) Full deviatoric moment tensor solutions (double-couple + CLVD) for 13 earthquakes (with \mbox{$M_l \geq 3.5$}), color-coded by depth, with beachball size scaled by moment magnitude. Green dots show relocated hypocenters from the full catalog (all events relocated by \texttt{GrowClust} plus 9 supplemented from \texttt{NonLinLoc}). (b–d) Rose diagrams showing the distributions of strike, dip, and rake for all 37 earthquakes, separated into the western (green) and eastern (orange) fault systems. (b) Strike distributions are consistent with the geometry of the two subparallel fault systems inferred from relocated seismicity. (c) Dip distributions indicate moderately to steeply dipping fault planes; dashed lines denote the mean dip orientations of the western and eastern fault systems. (d) Rake distributions are clustered around $\sim$ 270$^\circ$, indicating predominantly normal faulting with a minor strike-slip component.}
    \label{fig:focal_mt}
\end{figure}

To constrain the faulting style of the larger earthquakes within the swarm, we performed moment tensor (MT) inversions for 37 events ($M_l \geq 3.0$) using \texttt{MTTime} \cite{ichinose2014moment} (see SI materials for details, Text S3). Green's functions were computed with the CPS package \cite{herrmann2013computer} using the one-dimensional velocity model derived in this study. Relative locations from \texttt{GrowClust} were used where available, and \texttt{NonLinLoc} hypocenters were adopted for the remaining nine events. Waveforms were filtered between 0.05 and 1 Hz, and all selected solutions achieved variance reductions exceeding 50\% (SI Table~\ref{tab:mt_solution}).

The MT solutions indicate predominantly normal faulting with minor strike-slip components (Figures~\ref{fig:focal_mt} and \ref{fig:mtinv_example}). Fault strikes closely follow the north–south trends of the linear structures outlined by the relocated seismicity (Figure~\ref{fig:focal_mt}b), whereas dip angles cluster around $\sim30^\circ$ and $\sim60^\circ$, consistent with conjugate normal faults (Figure~\ref{fig:focal_mt}c). The rake angles are likewise consistent with dominantly extensional slip (Figure~\ref{fig:focal_mt}d). Together, these observations indicate reactivation of pre-existing extensional fault structures and justify interpreting the two relocated seismicity clusters as outlining two subparallel normal faults which we refer to as the western and eastern faults in the subsequent analysis. 

\begin{figure}[ht]
    \centering
    \includegraphics[width=1.0\linewidth]{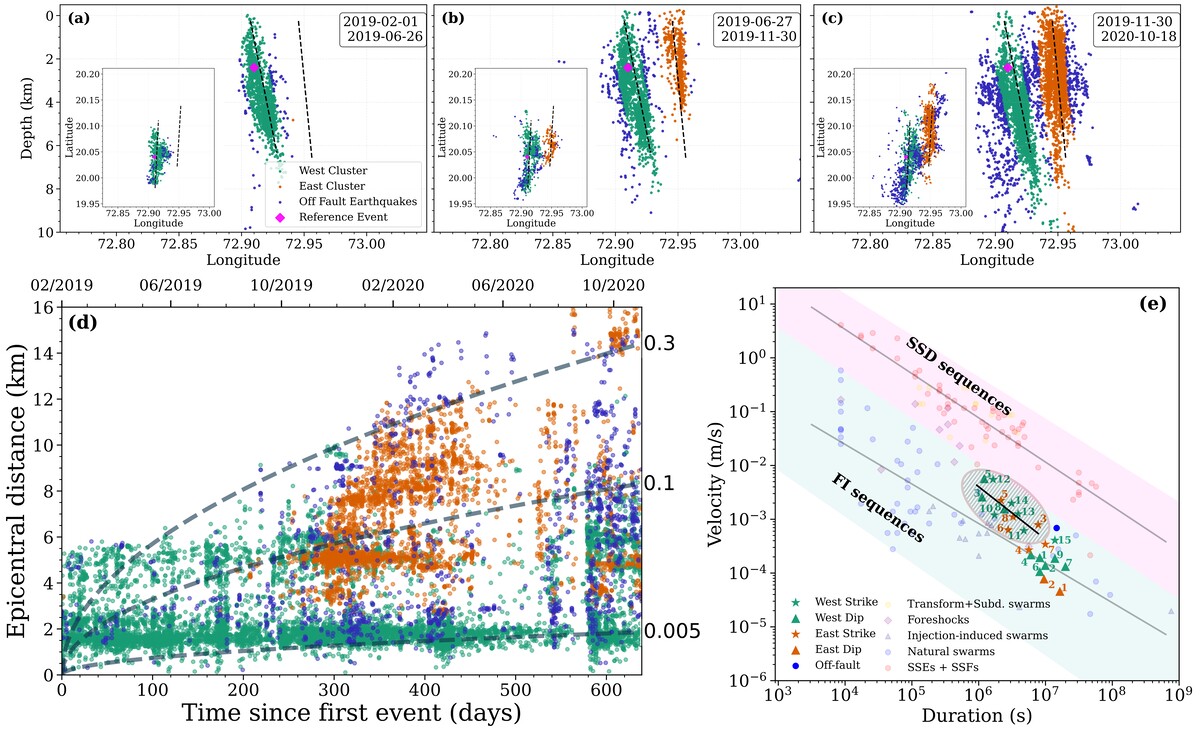}
    \caption{Spatiotemporal evolution and migration characteristics of the Palghar swarm. (a–c) Relocated earthquakes projected onto longitude–depth sections for the western (green), eastern (orange), and off-fault (blue) clusters. The pink diamond marks the first detected event, which defines the origin for the spatial and temporal evolution. Insets show map views. (d) Epicentral distance versus time with reference diffusion fronts (0.005, 0.1, and 0.3 m$^2$/s). (e) Migration velocity as a function of duration for representative migration episodes in the Palghar swarm compared with fluid-induced (FI) and slow-slip-driven (SSD) sequences from \citeA{danre2024parallel}. The blue and pink shaded areas arbitrarily illustrate the FI and SSD regimes, respectively, as in Figure~2 of \citeA{danre2024parallel}. Stars and triangles indicate along-strike and along-dip migration; green, orange, and blue symbols denote western-fault, eastern-fault, and off-fault migration. The hatched region marks the faster migration episodes used for the fit (black line) and slope--intercept analysis (SI Figure~\ref{fig:danre_stats}).
    }
    \label{fig:migration}
\end{figure}

\begin{figure}[ht]
    \centering
    \includegraphics[width=1.0\linewidth]{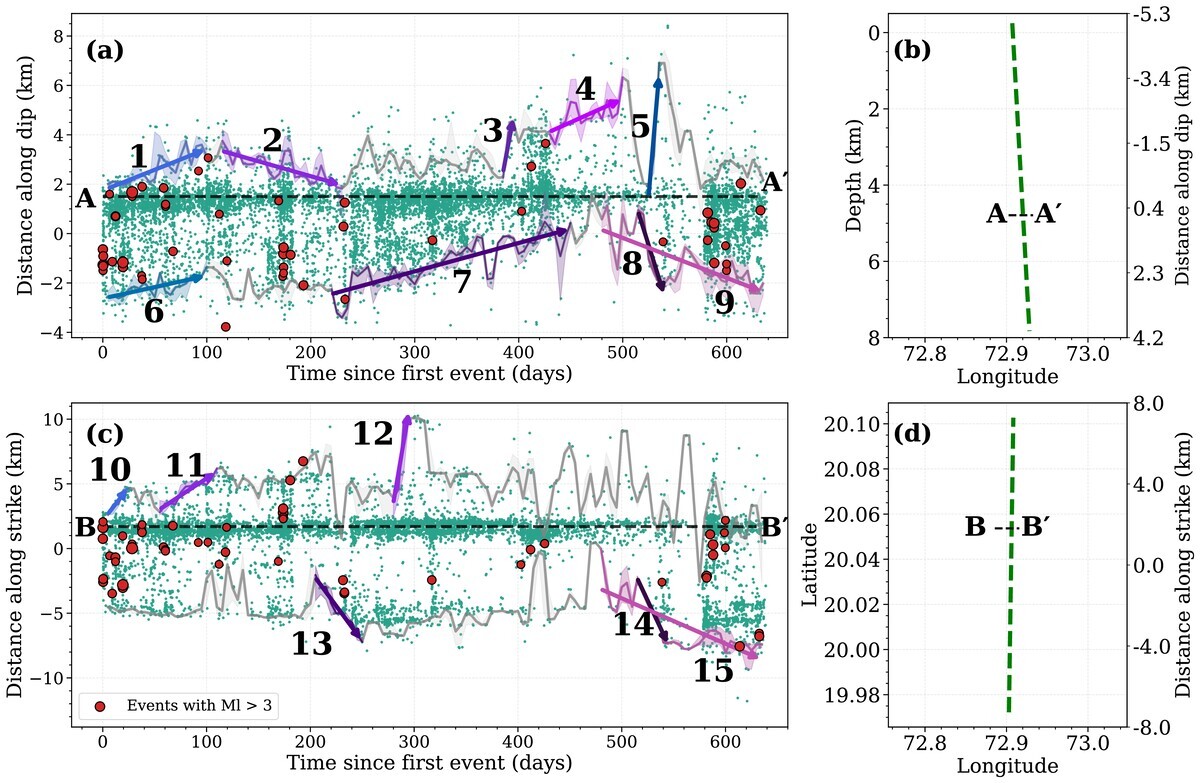}
    \caption{Spatiotemporal evolution of seismicity migration along the western fault. (a,c) Relocated earthquakes are projected onto along-dip and along-strike profiles, respectively, as a function of time since the first detected event. Green dots represent relocated earthquakes, and red circles denote events with $M_l$$>$  3. The gray envelope represents the 5th and 95th percentiles of seismicity. Colored arrows highlight individual migration fronts selected for linear regression to estimate migration velocities; arrow direction indicates the direction of propagation and the numbers identify the corresponding migration episodes. (b,d) Geometry of the along-dip and along-strike projection profiles used to calculate migration distances.}
    \label{fig:along_trace_distance_west}
\end{figure}

\begin{figure}[ht]
    \centering
    \includegraphics[width=1.0\linewidth]{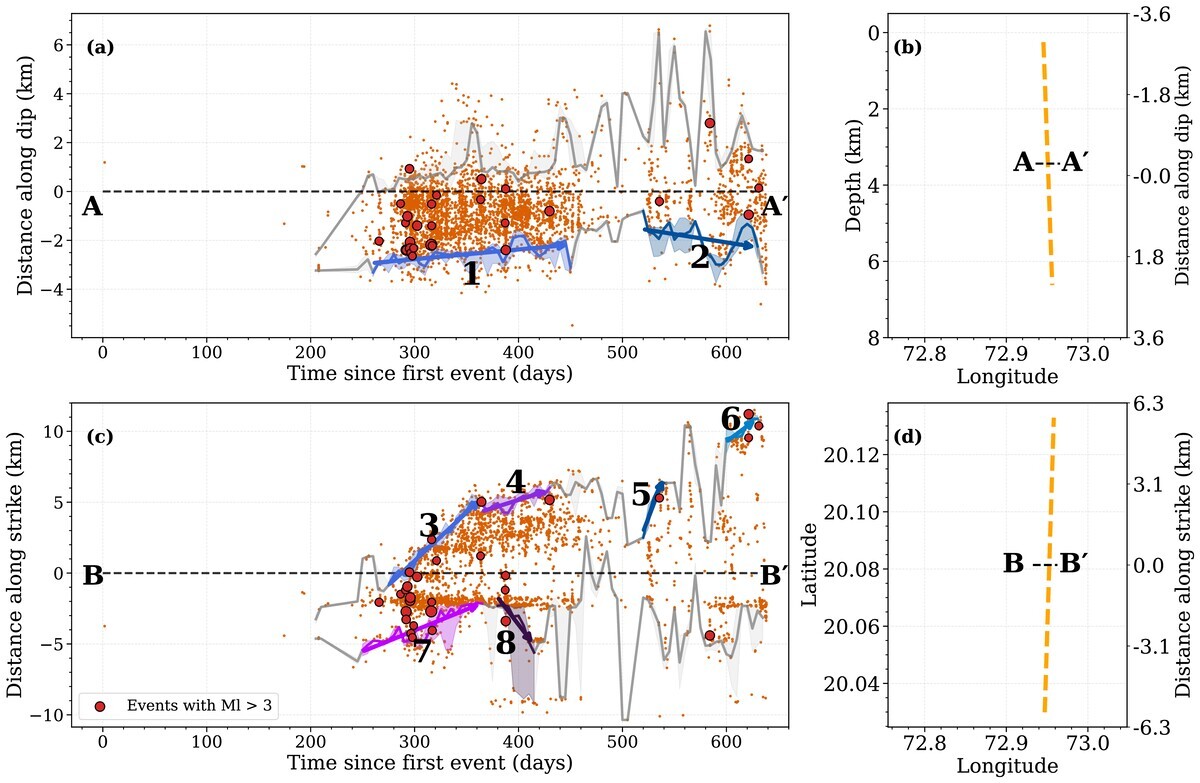}
   \caption{Spatiotemporal evolution of seismicity migration along the eastern fault. (a,c) Relocated earthquakes are projected onto along-dip and along-strike profiles, respectively, as a function of time since the first detected event. Orange dots represent relocated earthquakes, and red circles denote events with $M_l$$>$  3. The gray envelope shows the 5th and 95th percentiles of seismicity. Numbered arrows mark the major migration episodes identified along dip and strike, with arrow direction indicating the direction of propagation. (b,d) Geometry of the along-dip and along-strike projection profiles used to calculate migration distances.}
    \label{fig:along_trace_distance_east}
\end{figure}

\section{Results}
The high-resolution ML-enhanced relocated earthquake catalog developed in this study provides substantially improved constraints on the spatiotemporal evolution of the Palghar swarm compared to the human-picked catalogs that earlier studies have mostly relied upon \cite{sharma2020long, sharma2023characteristics, srinagesh2020appraisal} (SI Figure~\ref{fig:diffusion_same_color}). The improved spatial and temporal resolution reveals a systematic evolution of seismicity involving sequential activation of two subparallel fault systems and the progressive development of off-fault seismicity. During the first $\sim$200 days, seismicity remained confined to the western fault before abruptly nucleating on the eastern fault (Figures~\ref{fig:migration}a,b). Following the activation of the eastern fault, seismicity progressively expanded into the intervening region between the two faults, revealing possibly the progressive reactivation of an off-fault damage zone between the two fault systems. We classify earthquakes as western, eastern, or off-fault depending on whether they occur within a band of $\pm$1 km around the visually interpreted fault structures. The `off-fault' earthquakes predominantly activated only after seismicity began to cluster around both faults, suggesting that ruptures were triggered mechanically by elastic stress perturbations generated by the simultaneous activation of slip on both faults.

We characterize the temporal evolution of the swarm using the epicentral distance-time distribution of the relocated earthquakes, where epicentral distance is measured relative to the first event detected in the machine-learning-enhanced catalog (origin time: 01/02/2019 00:01:53; latitude: $20.04^\circ$; longitude: $72.91^\circ$; depth: 2.4 km) (Figure~\ref{fig:migration}d). Seismicity expands progressively with time and is broadly bounded by diffusion fronts corresponding to diffusivities of approximately 0.005--0.3~m$^{2}$~s$^{-1}$ (Figure~\ref{fig:migration}d). However, this overall apparently diffusive expansion is, in reality, a superposition of distinct propagating fronts, many of which are faster than diffusive, and seismicity patterns that outline different clusters of seismicity that lie either on the two faults or within the region bounded by them (Figure~\ref{fig:migration}a-c). The propagating fronts define the leading edge of the swarm and are most pronounced during the activation of the eastern fault between $\sim$260 and 460 days, when earthquakes progressively migrate to larger epicentral distances. In contrast, the western fault remains active throughout most of the sequence and repeatedly generates migration fronts that appear to originate from a persistent source region located approximately 1.5 km from the epicenter of the reference event.

The inferred `fault' structures shown in Figures~\ref{fig:migration}a-c clearly indicate that the epicentral distance migration is nearly insensitive to any along-dip migration, given the steeply dipping normal `faults' inferred from the seismicity clustering. Given this, we resolve separately the along-strike and dip migrations of the western and eastern clusters assuming the linear structures inferred in Figures~\ref{fig:migration}a-c to better constrain the processes driving the swarm (Figures~\ref{fig:along_trace_distance_west} and \ref{fig:along_trace_distance_east}). The western fault exhibits two clear end-member seismicity clustering patterns -- 1. a persistent, spatially confined seismicity band at about $\sim$5 km depth that survives throughout the duration of the swarm, and 2. repeated, episodic migration fronts that seem to emanate from this persistent band and propagate both along strike and dip (Figure~\ref{fig:along_trace_distance_west}). These fronts migrate both away from and back toward the persistent seismicity band, producing multiple outward- and back-propagating migration episodes throughout the sequence. 

In contrast, the eastern fault displays a distinctly different migration style (Figure~\ref{fig:along_trace_distance_east}). Following its activation, migration in this cluster is dominated by rapid northward propagation along strike, accompanied by slower down-dip propagation of the seismicity front. Unlike its western counterpart, the eastern cluster lacks a persistent seismicity band from which these migration episodes originate. A transient along-strike seismicity band does develop but gradually disappears after approximately 450 days. Migration velocities estimated from representative episodes marked in both Figures~\ref{fig:along_trace_distance_west} and ~\ref{fig:along_trace_distance_east} on both fault systems are compared with the velocity-duration classification of \citeA{danre2024parallel} in Figure~\ref{fig:migration}e.

The off-fault seismicity exhibits migration patterns that closely mirror those observed on the principal fault systems. The off-fault cluster becomes most prominent only after the eastern fault also activates at around 260 days. Between approximately 260 and 360 days, the epicentral migration front of the off-fault earthquakes appears remarkably coordinated with that of the eastern fault cluster (SI Figure~\ref{fig:off_fault_500_m}). Again, during the later stages of the swarm (approximately 480--640 days), off-fault seismicity clustered around the western fault similarly reproduces the southward and up-dip migration observed along the western fault (SI Figure~\ref{fig:off_fault_migration}). These relationships remain unchanged when the definition of the fault-zone width is reduced from $\pm$1 km to $\pm$500 m, demonstrating that the observed coordinated migration patterns are robust. The close correspondence between the on-fault and off-fault migrations indicates that deformation evolved mechanically coherently across the entire fault network rather than remaining confined to individual fault planes.

\section{Discussion}

The new high-resolution earthquake catalog reveals a substantially more complex evolution of the Palghar swarm than previously recognized \cite{srinagesh2020appraisal,sharma2020long,sharma2023characteristics,nath2021dynamic}. Rather than occurring on a single fault zone, the swarm initiated on one of two closely spaced fault segments before migrating to the second one and, finally, progressively expanded into the surrounding fault damage zone. The relocated seismicity further reveals persistent fault-localized activity on the western fault, together with transient propagating fronts, indicating that the observed swarm evolution cannot be explained by a single triggering mechanism. These newly resolved spatiotemporal patterns provide important constraints on the physical processes governing the Palghar swarm and earthquake swarms in general.

The overall migration of seismicity appears to follow a broad diffusive envelope, with apparent hydraulic diffusivities ranging from approximately 0.005 to 0.3~m$^{2}$~s$^{-1}$, comparable to values reported for fluid-driven earthquake swarms elsewhere \cite{goebel20172016,cabrera2021tracking}. However, this apparent diffusion masks the underlying propagation dynamics, which are only revealed by resolving the geometry and evolution of the western, eastern, and intervening off-fault seismicity clusters. The resulting high-resolution view reveals distinct, coherent migration fronts within and between these clusters, several of which propagate substantially faster than expected from fluid diffusion alone. This also shows that, in the absence of high resolution migration data, diagnosing mechanistic processes driving earthquake swarms based on overall migration patterns can be misleading.

\citeA{danre2024parallel} classify swarms as fluid- and slow-slip driven, based primarily on the slope and intercept of their migration velocity-duration scaling on log-log axes. In the velocity-duration space of \citeA{danre2024parallel} (Figure~\ref{fig:migration}e), the migration episodes of the Palghar swarm identified in Figures~\ref{fig:along_trace_distance_west} and \ref{fig:along_trace_distance_east} distribute nearly equally within two clusters that exhibit different scalings. We include episodes 3, 5, 6, and 8 from the eastern; and 3, 5, 8, 10, 11, 12, and 14 from the western faults within the hatched cluster in Figure~\ref{fig:migration}e, while the rest of the migration episodes from the two faults appear to cluster separately. The slope and intercept of the velocity--duration scaling represented by the hatched cluster are both systematically larger than the fluid-driven regime recognized by \citeA{danre2024parallel} in their Figure~3a and b (SI Figure~\ref{fig:danre_stats}). The slope, in particular, is even larger than the slow-slip-driven regime inferred by \citeA{danre2024parallel} while the intercept lies squarely within it. The unhatched cluster of migration episodes in Figure~\ref{fig:migration}e, in contrast, is distributed predominantly along the fluid-driven scaling reported by \citeA{danre2024parallel}. It is also interesting that the hatched cluster is composed predominantly of along-strike migrations, while the `fluid-driven' cluster is predominantly composed of along-dip migrations. This is consistent with the geological intuition that fluids predominantly travel along the dip of a fault, exploiting the largest pressure gradients along permeable pathways. However, it is worth noting that the \citeA{danre2024parallel} classification applies strictly to epicentral migrations, with the slow-slip-driven regime almost entirely sampled from plate-boundary settings. Nevertheless, the two clusters of migration episodes in Figure~\ref{fig:migration}e, the differences in the velocity-duration scaling between them, and their simultaneous occurrence throughout the evolution of the swarm together provide a consistent indication that fluid diffusion alone cannot explain the swarm's evolution. Instead, multiple transient processes likely associated with coupled fluid-migration and aseismic slip contributed to earthquake migration.

The inference that multiple processes contributed to swarm evolution is also consistent with contrasting migration styles on the western and eastern fault systems. The western fault hosts a persistent, spatially confined seismicity band from which seismicity repeatedly expands and contracts along strike and dip (Figure~\ref{fig:along_trace_distance_west} a, c). In contrast, the eastern fault lacks a comparable persistent seismicity cluster and is instead dominated by a transient along-strike migration episode that contains several of the M$_l>$3.0 earthquakes (Figure~\ref{fig:along_trace_distance_east}a, c). The distinct behavior of the two fault systems, despite their close spatial proximity and similar orientation, suggests that they responded differently to the evolving stress field and fluid-pressure conditions during swarm evolution.

An equally important observation is that the migration of seismicity within the distributed fault volume is temporally and kinematically coupled to migration on the principal faults. Such closely coordinated migration of the on- and off-fault seismicity likely indicates mechanical interactions between the fault and its surrounding volume through elastic stress transfer. For example, the 260--460 day coordinated migration episode on the eastern fault falls within the slow-slip-associated regime in the velocity–duration scaling of \citeA{danre2024parallel}, suggesting that aseismic deformation may have contributed to stress redistribution during this episode.

The depth distribution of seismicity provides additional constraints on the origin of the swarm. Earthquakes are strongly confined to depths shallower than approximately 6 km (Figure~\ref{fig:depth_section}c). Such a sharp lower bound may reflect structural termination of the active fault network, a permeability barrier that restricts deeper fluid migration, or a rheological transition at depth. Because the swarm occurs within the granitoid basement beneath the Deccan Volcanic Province, where extensive dyke-related fault networks are expected \cite{courtillot1986deccan,rubin1988dike,rubin1992dike}, structural termination at only $\sim$6 km appears unlikely \cite{rai2025deccan}. Likewise, a transition from velocity-weakening to velocity-strengthening friction is difficult to reconcile with the shallow depth cutoff, regional geothermal gradients \cite{gupta1984surface, Scholz1998, roy2000heat, scholz2019mechanics}, and evidence for deeper crustal low-velocity zones beneath nearby swarm regions \cite{rai2025deccan}. The persistent seismicity band on the western fault is itself confined to a region only about 1 km wide along both strike and dip, suggesting a highly localized source region. Taken together, we interpret the persistent seismicity band as being due to fluid circulation within a $\sim$5 km deep fluid-rich zone, while downdip fluid migration is restricted by permeability barriers or strong pore-pressure gradients. Periodic fluid overpressures within this zone could repeatedly perturb the surrounding fault, providing the initial driving mechanism for swarm nucleation. The intermittent migration episodes emanating from this seismicity band could then be associated with a fault-valving type mechanism where their velocity-duration regimes are determined by whether the propagation fronts are fluid- or aseismic-slip dominated \cite{sibson1992fault, zhu2020fault}. 

Previous studies have attributed the Palghar swarm, and similar earthquake swarms within the Deccan Volcanic Province, primarily to monsoon-driven groundwater recharge \cite{hainzl2014monsoon,gupta2018reservoir,sharma2020long,nath2021dynamic,gahalaut2022long,sharma2023characteristics}. The densified earthquake catalog, however, does not reveal a systematic relationship between rainfall and seismicity during the early and most productive phase of the swarm, although one increase in seismicity during the later stages follows the end of the monsoon season (SI Figure~\ref{fig:events_time_series}). While these observations do not exclude a contribution from meteoric recharge, they provide little evidence that rainfall alone controlled swarm initiation or sustained the subsequent evolution of the sequence. Instead, the relocated seismicity favors a model in which a deeper fluid source couples with the fault geometry and elastic stress redistribution to determine the evolution of the Palghar swarm.  

\section{Conclusion}
The machine-learning-enhanced, high-resolution relocated earthquake catalog presented in this study provides new insights into the physical mechanisms governing the evolution of the Palghar earthquake swarm. Rather than occurring on a single fault zone, the swarm evolved through the sequential activation of two closely spaced subparallel fault systems, followed by the progressive development of distributed off-fault seismicity within the intervening damage zone. The contrasting migration styles observed on the two fault systems, together with the coordinated evolution of the surrounding off-fault seismicity, reveal a substantially more complex deformation process than previously recognized.

The overall migration of seismicity follows a broad diffusive envelope, but our observations demonstrate that such behavior alone does not uniquely indicate fluid migration. Instead, the apparently diffusive expansion masks multiple migration episodes spanning distinct regimes in the velocity–duration space, indicating the simultaneous and coordinated influence of fluid migration and aseismic-deformation-mediated stress redistribution. This inference is further supported by the delayed activation of the eastern fault and the coordinated evolution of on- and off-fault seismicity. Our results show that the Palghar swarm does not behave as a sequence governed by a single end-member process; instead, fluid-assisted deformation and transient stress transfer couple within an interacting fault network, producing migration behavior that spans the regimes commonly associated with fluid diffusion and aseismic slip. More fundamentally, they suggest that the interplay among multiple transient processes may be particularly evident in intraplate settings, where low background tectonic strain rates enhance the sensitivity of fault systems to transient strength and stress perturbations. Our results underscore the unique value of intraplate earthquake swarms as natural laboratories for investigating the mechanics of earthquake triggering and fault interaction and motivate more systematic investigations of these systems.

\section*{Data Availability Statement}

No new seismic waveform data were collected for this study. The continuous waveform data used in this work were obtained from the seismic networks operated by the National Geophysical Research Institute (NGRI) and the National Center for Seismology (NCS), Ministry of Earth Sciences, Government of India, and were previously used by \citeA{sharma2020long, gahalaut2022long}. The machine-learning earthquake catalog developed in this study is available at \url{10.5281/zenodo.21219983}.
This study made use of the publicly available software packages PyVelest (\url{https://github.com/saeedsltm/PyVelest}), BPMF (\url{https://github.com/ebeauce/Seismic_BPMF}),
PhaseNet (\url{https://github.com/AI4EPS/PhaseNet}), NonLinLoc (\url{https://github.com/ut-beg-texnet/NonLinLoc}),
GrowClust (\url{https://github.com/dttrugman/GrowClust}), and MTTime (\url{https://github.com/LLNL/mttime}).
and Computer Programs in Seismology (CPS) (\url{https://rbherrmann.github.io/ComputerProgramsSeismology/index.html}).
The rainfall data used in the Supporting Information were obtained from the NASA Langley Research Center Prediction Of Worldwide Energy Resources (POWER) Project Daily API and are publicly available at \url{https://power.larc.nasa.gov/}.

\section*{Acknowledgments}

We thank the National Geophysical Research Institute (NGRI), Hyderabad, and the National Center for Seismology (NCS) for providing the continuous seismic waveform data used in this study. We thank L\'eonard Seydoux and Farzaneh Mohammadi for their assistance with implementing the BPMF workflow and developing the machine-learning earthquake catalog. We also acknowledge the developers of the open-source software packages BPMF, PhaseNet, NonLinLoc, GrowClust, MTTime, and Computer Programs in Seismology (CPS), which were used in different stages of the analysis. This work was supported by the Ministry of Education, Government of India, through the Scheme for Promotion of Academic and Research Collaboration (SPARC) project (Project No. SPARC/2019--2020/P1928/SL). Computational resources were provided by the Byerlee and Rice high-performance computing clusters at the School of Earth and Planetary Sciences, National Institute of Science Education and Research (NISER), Bhubaneswar. HSB also acknowledge partial support from the ERC Consolidator grant PERSISMO (\#865411). 

\clearpage

\section*{Supporting Information}
\renewcommand{\thefigure}{S\arabic{figure}}
\setcounter{figure}{0}

\begin{figure}[ht]
    \centering
    \includegraphics[width=0.85\linewidth]{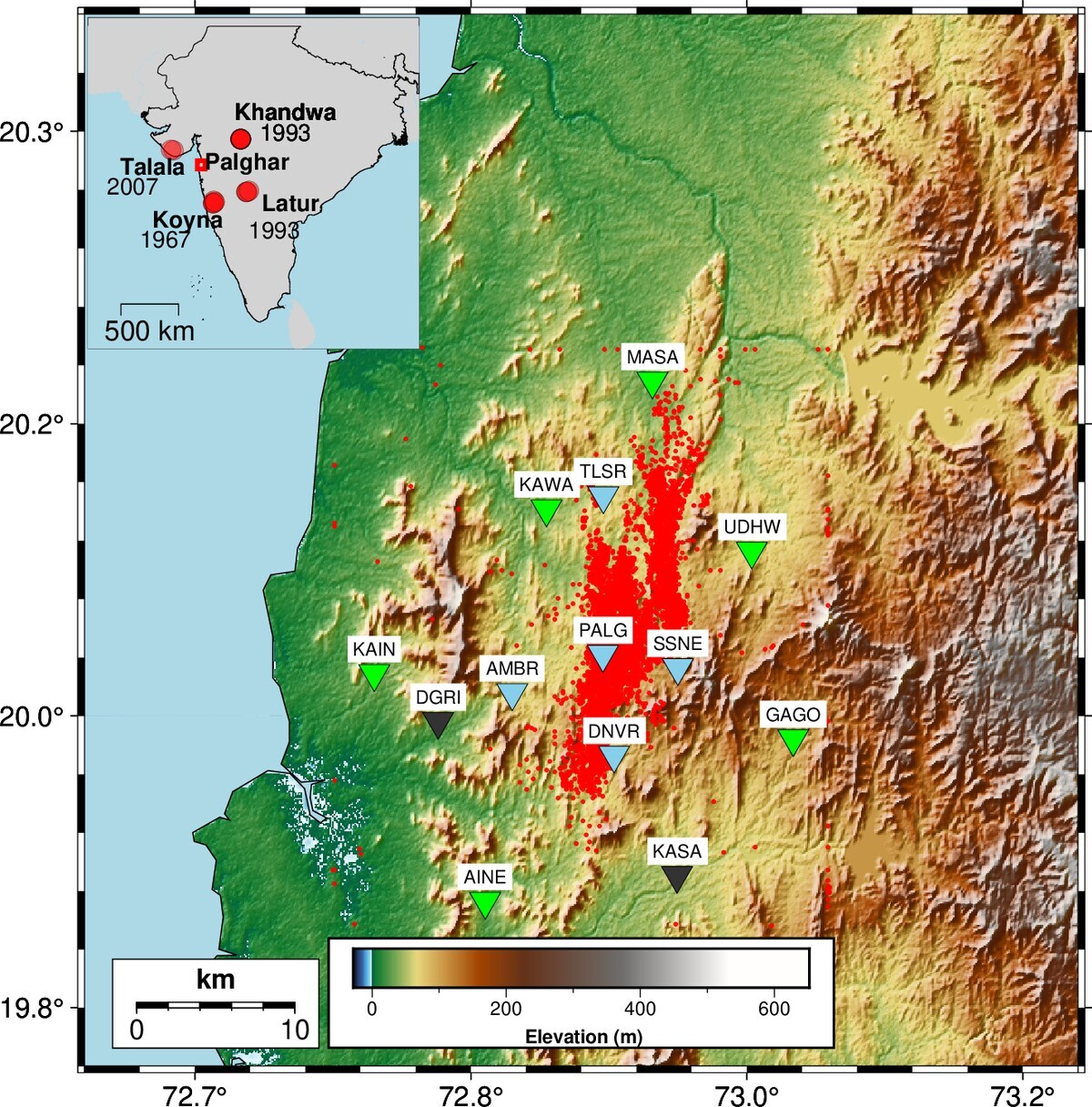}
    \caption{Map of the Palghar region showing the distribution of the seismic stations (triangles) used in this study. Green triangles denote stations operated by NGRI, while sky-blue triangles denote those operated by NCS. Note that the KASA and DGRI stations were later relocated to TLSR and DNVR, respectively, which were also managed by NCS. Red dots represent earthquake epicenters derived from machine learning–based detections and located using NonLinLoc. The inset map of India highlights the Palghar region (red box) and marks other major intraplate earthquakes and swarm occurrences across the Indian peninsular region.}
    \label{fig:intro_map}
\end{figure}

\section*{Text S1. Velocity Model Constraints and Robustness Tests}\label{tex:si_velo}
To evaluate the robustness of the final 1-D velocity model derived for the Palghar swarm, we examine the stability of the inversion results, the sensitivity of the recovered velocity structure, and the consistency of the adopted $V_p/V_s$ ratio.
\subsection*{S1.1 Estimation of the $V_p/V_s$ Ratio}

The $V_p/V_s$ ratio used in the velocity inversion was independently estimated from the manually picked P- and S-wave arrival times of the earthquakes used for velocity modeling.
\begin{figure}[ht]
    \centering
    \includegraphics[width=0.8\linewidth]{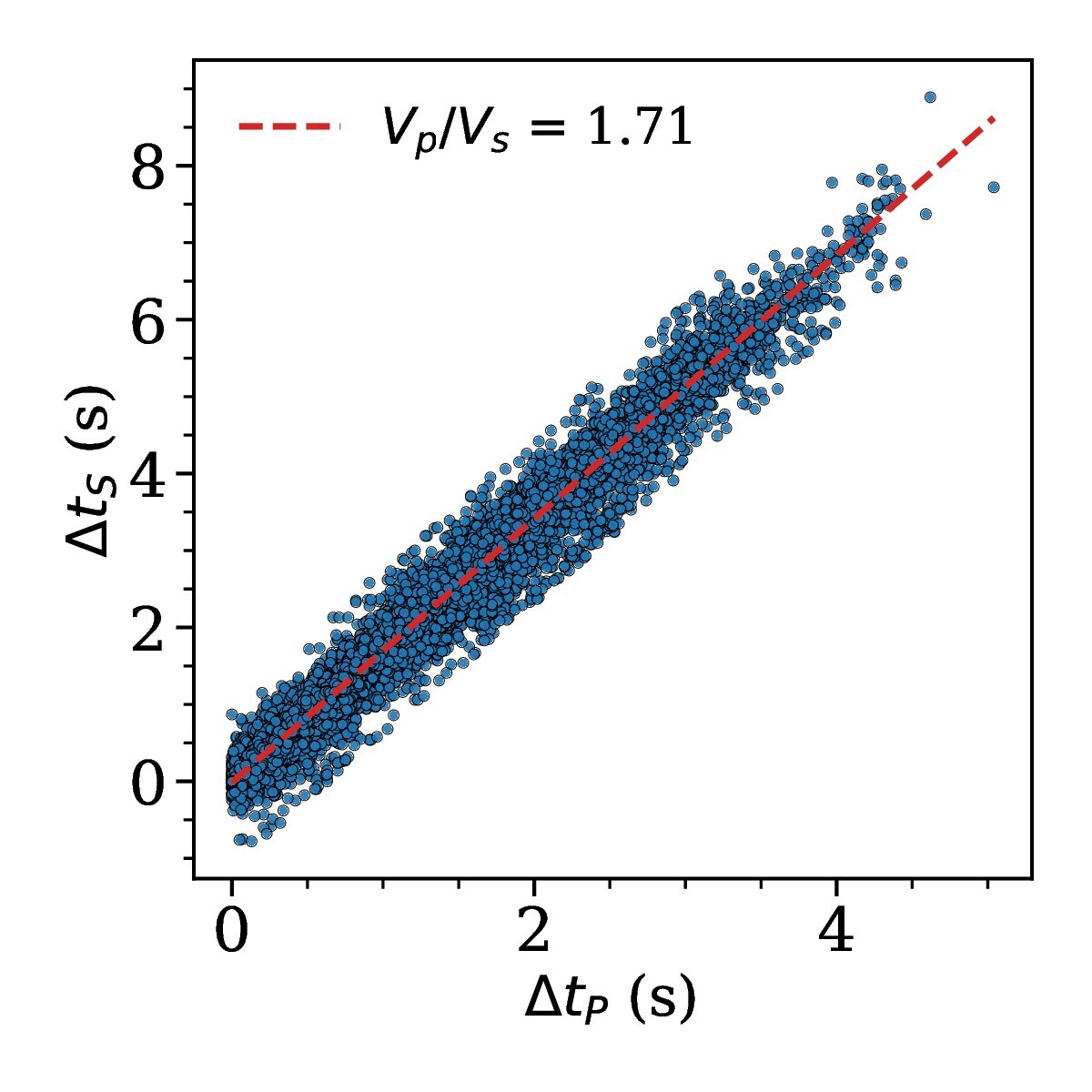}
    \caption{Relationship between the differential S- and P-wave travel times ($\Delta t_S$ and $\Delta t_P$) for the earthquakes used to estimate the $V_p/V_s$ ratio. The dashed red line represents the best-fitting linear regression, yielding $V_p/V_s \approx 1.71$.}
    \label{fig:vp_vs}
\end{figure}
Figure {\ref{fig:vp_vs} shows the relationship between the differential S- and P-wave travel ($\Delta t_S$ and $\Delta t_P$) times measured at the same stations for the selected earthquakes. The slope of the best-fitting regression line yields a $V_p/V_s$ ratio of approximately $1.71$. 

\subsection*{S1.2 RMS Residual Statistics of Velocity Inversions}
To reduce dependence on the starting velocity model, synthetic velocity models were generated by perturbing the reference model of \citeA{srinagesh2020appraisal}. Layer velocities were randomly perturbed within $\pm0.7$ km s$^{-1}$ and interface depths within $\pm2$ km to produce 70 alternative starting models for each inversion stage. Each model was inverted independently using \texttt{VELEST} in simultaneous inversion mode. We adopted \texttt{INVERTRATIO = 2}, such that the velocity model was updated only every second iteration, while earthquake hypocenters and station corrections were updated during every iteration. This strategy allows the hypocenters and station delays to stabilize before subsequent velocity updates, thereby improving inversion stability and convergence. After the completion of each stage, consisting of 7 iterations, the final velocity model was obtained by computing the layer-by-layer mean of all successfully inverted velocity models. This averaged model was then used as the starting model for the next inversion stage and was again perturbed randomly to generate an ensemble of starting models. Three successive inversion stages were performed, resulting in a total of 210 inverted velocity models, from which the final mean velocity model was selected for use in all subsequent analyses.
\begin{figure}[ht]
\centering
\includegraphics[width=0.8\linewidth]{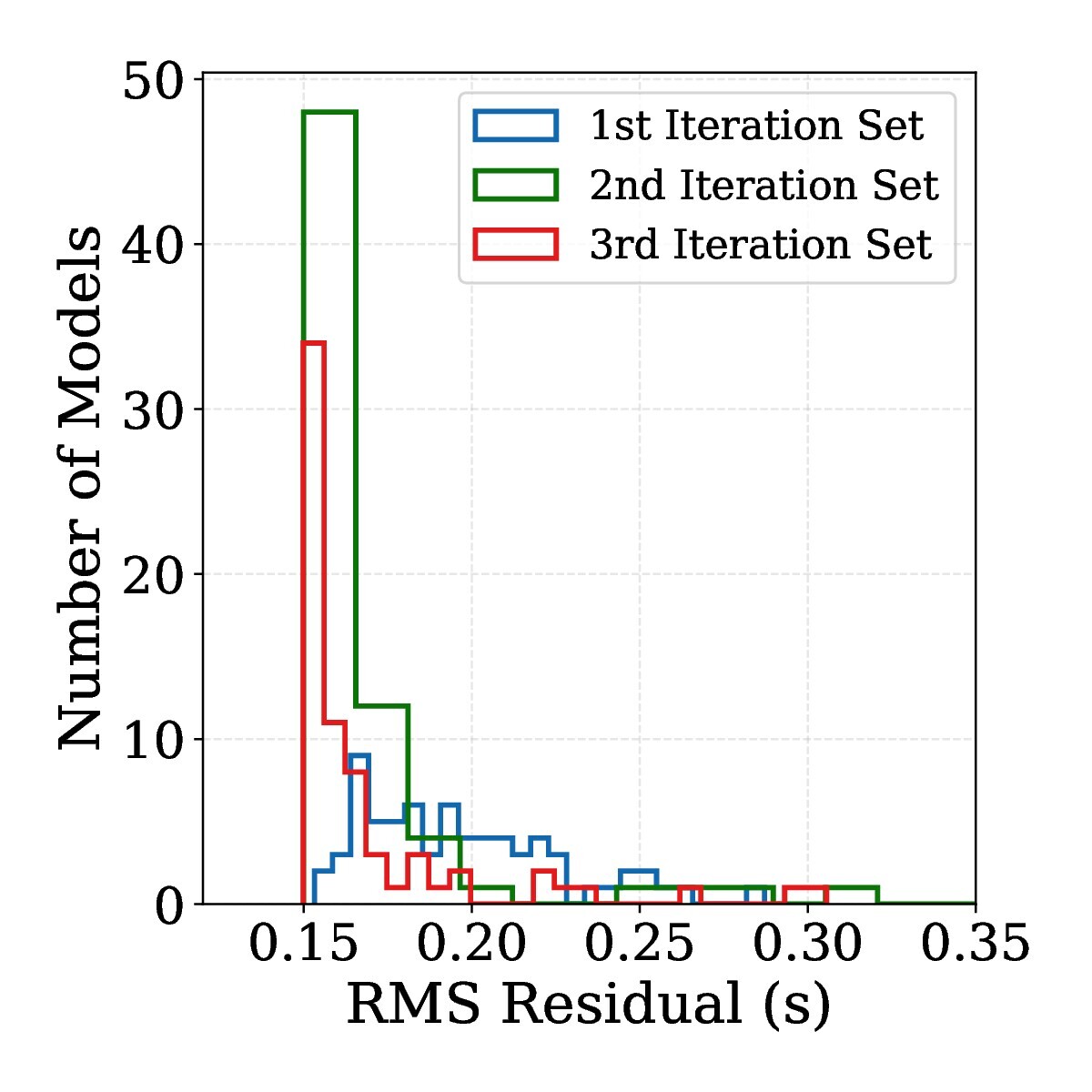}
\caption{Distribution of RMS travel-time residuals obtained from velocity inversions during the three iteration stages of the inversion. The progressive reduction and convergence of RMS values indicate that the inversion approaches a stable velocity model.}
\label{fig:supp_rms}
\end{figure}

Figure~\ref{fig:supp_rms} shows the distribution of RMS travel-time residuals obtained from the inversions during the three iteration stages described in the main text. The first iteration set displays a broader range of RMS values reflecting the diversity of starting models. With subsequent inversion stages, the RMS residuals progressively decrease and converge toward a narrow range of values, indicating that the inversion approaches a stable minimum-misfit solution.

Using the final velocity model, earthquakes were relocated with \texttt{HypoInverse}. A comparison of location error ellipses before and after velocity inversion is shown in Figure~\ref{fig:error_ellipse}, illustrating the reduction in location uncertainties achieved with the updated model.

\begin{figure}[ht]
  \centering
  \includegraphics[width=1.0\textwidth]{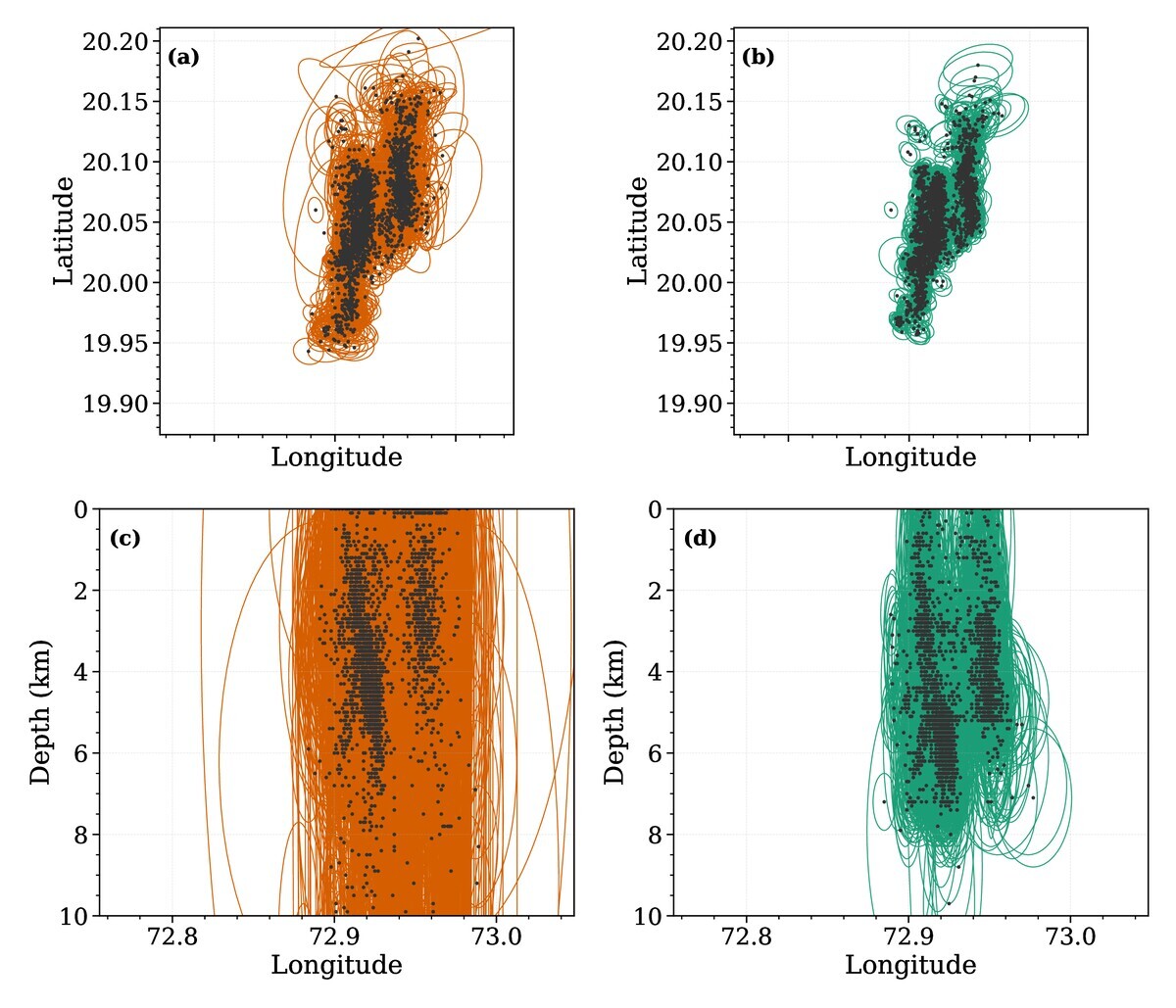}
  \caption{Comparison of earthquake location uncertainty ellipses before and after velocity inversion. Panels (a) and (c) show epicentral and depth–longitude distributions obtained using the initial velocity model, while panels (b) and (d) show results after relocation using the final velocity model derived in this study. Ellipses represent location uncertainties estimated from \texttt{HypoInverse}. The comparison illustrates the reduction in location uncertainty achieved using the updated velocity model.}
  \label{fig:error_ellipse}
\end{figure}
\begin{figure}[ht]
    \centering
    \includegraphics[width=1.0\linewidth]{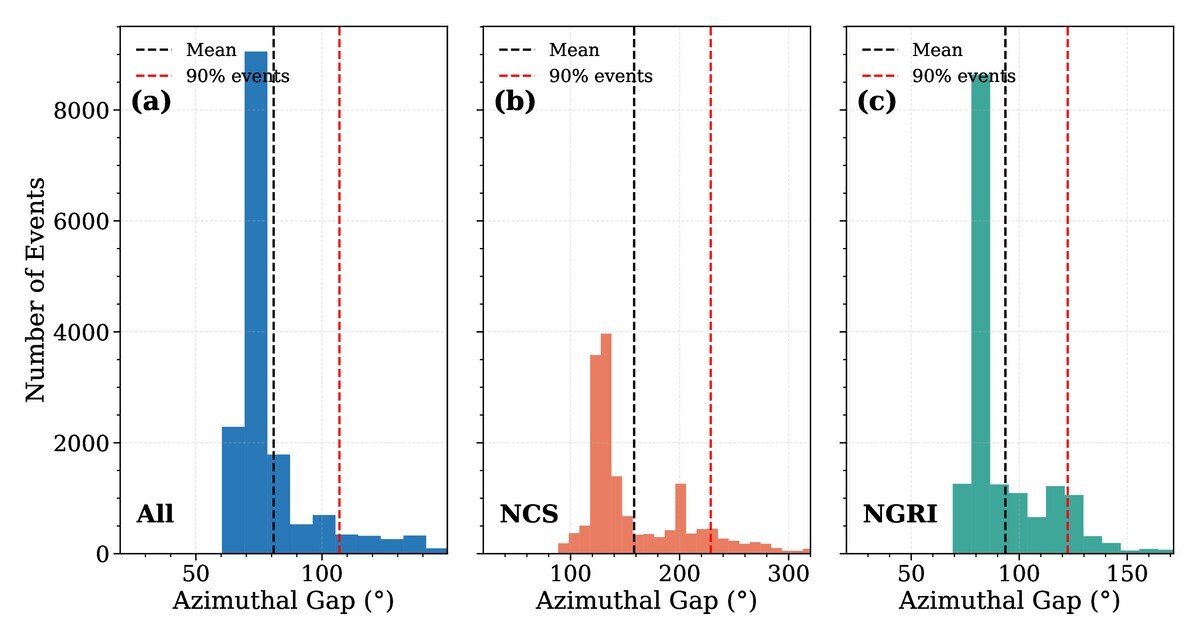}
    \caption{Distribution of azimuthal gap values for earthquake locations obtained using different seismic network configurations. (a) Azimuthal gap distribution for all relocated events used in this study. (b) Azimuthal gap distribution for locations determined using only the NCS network. (c) Azimuthal gap distribution for locations determined using only the NGRI network. Black dashed lines indicate the mean azimuthal gap for each dataset, while red dashed lines mark the azimuthal gap below which 90\% of the events are located. The distributions show that most relocated events have azimuthal gaps below $\sim$107$^\circ$ for the full dataset, whereas the individual NCS and NGRI network solutions exhibit larger azimuthal gaps due to reduced station coverage.}
    \label{fig:gap_comparison}
\end{figure}

\begin{table}[h]
\centering
\caption{Seismic station information used in this study. Latitude and longitude are given in degrees, and elevation is given in meters above sea level.}
\begin{tabular}{lccc}
\hline
Station & Latitude (°N) & Longitude (°E) & Elevation (m) \\
\hline
AINE & 19.8940 & 72.8420 & 73 \\
GAGO & 19.9870 & 73.0280 & 101 \\
KAIN & 20.0240 & 72.7750 & 37 \\
KAWA & 20.1180 & 72.8790 & 51 \\
MASA & 20.1910 & 72.9430 & 44 \\
UDHW & 20.0940 & 73.0030 & 57 \\
PALG & 20.0350 & 72.9131 & 162 \\
TLSR & 20.1255 & 72.9133 & 57 \\
SSNE & 20.0275 & 72.9583 & 130 \\
AMBR & 20.0131 & 72.8583 & 70 \\
DNVR & 19.9775 & 72.9198 & 78 \\
\hline
\end{tabular}\label{tab:station_info}
\end{table}

\section*{Text S2. Earthquake detection, phase picking, and relocation}

Earthquake detection and initial earthquake locations were obtained using the BackProjection and Matched-Filtering (BPMF) workflow \cite{beauce2024bpmf}. Within the workflow, continuous three-component waveform data from the NGRI and NCS seismic networks were corrected for instrument response and band-pass filtered between 1 and 20 Hz. The original pretrained \texttt{PhaseNet} model \cite{zhu2019phasenet} was applied to the continuous waveforms to generate continuous P- and S-wave probability time series, which were subsequently backprojected across the seismic network to identify coherent earthquake signals. Earthquake detections were identified using a backprojection threshold of 2 (SI Figure~\ref{fig:beam_power}), producing an initial machine-learning (ML) catalog of 57,826 detected events. The ML catalog recovered approximately 95\% of the manually picked high-quality earthquakes (recorded by at least six stations) while substantially increasing the total number of detected events.

For each detected event, P- and S-wave arrival times were refined using \texttt{PhaseNet} in picking mode. The automatically determined phase arrivals were subsequently used as input to the probabilistic earthquake location program \texttt{NonLinLoc} \cite{lomax2000probabilistic}. Earthquake hypocenters were determined using the Equal Differential Time (EDT\_OT\_WT\_ML) location algorithm, which combines differential travel-time residuals, weighted phase observations, and maximum-likelihood estimation to obtain robust hypocenter locations that are less sensitive to outlying phase picks. Only events with picks from at least six stations were considered for location, resulting in 20,638 well-located earthquakes. Approximately 88\% of the manually picked earthquakes were successfully recovered after the initial location stage.

The \texttt{NonLinLoc} catalog was subsequently relocated using \texttt{GrowClust} \cite{trugman2017growclust}, which refines relative earthquake locations by inverting differential travel times derived from waveform cross-correlation. Relative relocations were performed for event pairs with waveform cross-correlation coefficients of at least 0.8. This procedure yielded a final relocated catalog of 16,008 earthquakes, corresponding to approximately 81\% of the manually picked catalog. The relocated catalog forms the basis for all spatiotemporal analyses presented in this study.

Uncertainties in earthquake locations were estimated using the GrowClust bootstrap procedure, in which hypocenter locations are perturbed, and the relocation is repeated to evaluate solution stability. Horizontal and vertical location uncertainties for events relocated with \texttt{GrowClust} are shown in SI Figure~\ref{fig:NC_GC}, which compares these uncertainties with those obtained from \texttt{NonLinLoc}, where location errors are derived from confidence regions of the posterior probability density function. The comparison highlights the improved relative location precision achieved using the cross-correlation-based relocation approach.

\begin{figure}[ht]
    \centering
    \includegraphics[width=1.0\linewidth]{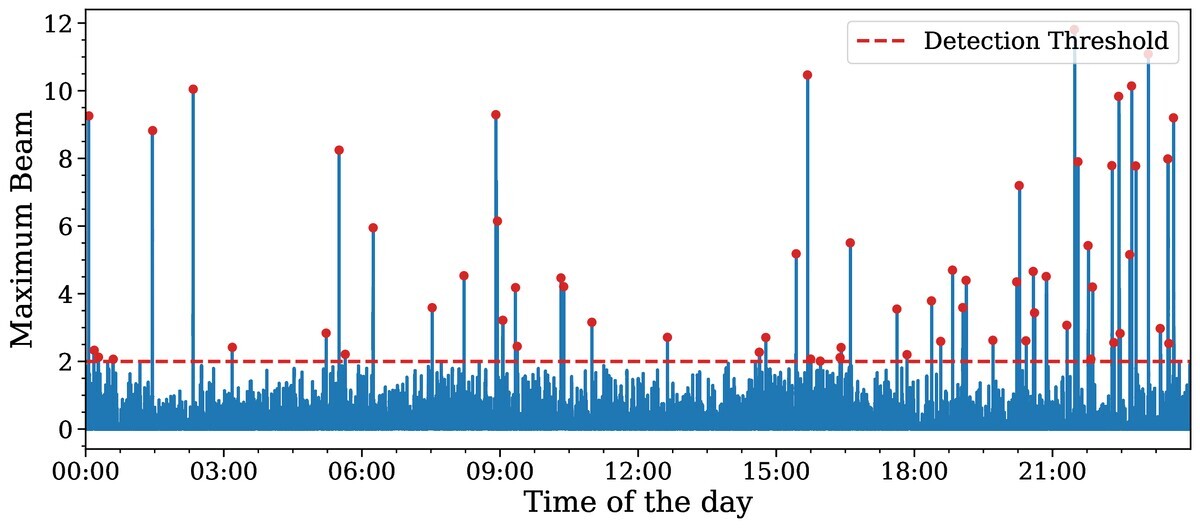}
    \caption{Example of the maximum beam power time series computed across the seismic network for a single day during the swarm sequence. The red dashed line marks the detection threshold (beam power = 2), and peaks exceeding this threshold were identified as candidate earthquakes. A total of 63 detections were identified for the day shown.}
    \label{fig:beam_power}
\end{figure}
\begin{figure}[ht]
    \centering
    \includegraphics[width=1.0\linewidth]{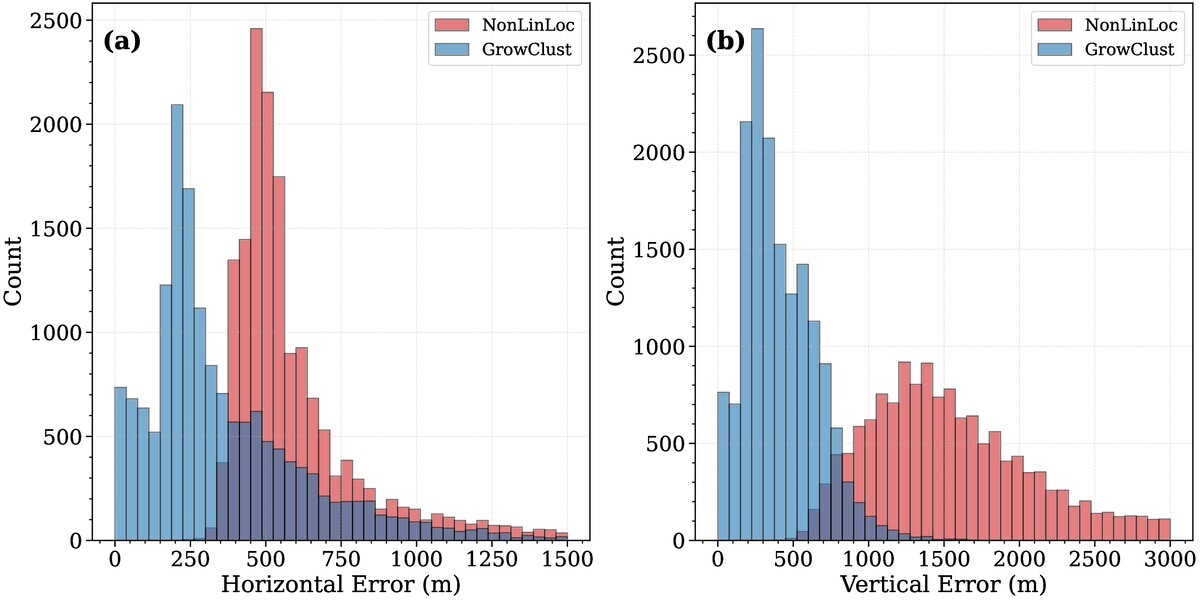}
    \caption{Comparison of earthquake location uncertainties obtained from \texttt{NonLinLoc} and \texttt{GrowClust}, showing (a) horizontal and (b) depth uncertainties.}
    \label{fig:NC_GC}
\end{figure}

\section*{Text S3. Moment tensor solutions}

This section provides additional details of the moment tensor inversion procedure summarized in the main text. Moment tensor inversions were performed using \texttt{MTTime} \cite{ichinose2014moment}, with displacement Green's functions computed using the Computer Programs in Seismology (CPS) package \cite{herrmann2013computer} for the final one-dimensional velocity model derived in this study. Instrument responses were removed prior to processing, and the three-component broadband waveforms were converted to displacement and band-pass filtered between 0.05 and 1 Hz using a two-pole Butterworth filter. For each station, waveform windows were extracted around the P-wave arrival, extending from 2 s before the arrival to 6--8 s after the arrival, with the post-arrival window length adjusted according to the source--receiver distance.

Inversion quality was evaluated using the variance reduction (VR) between the observed and synthetic waveforms across all available stations. Only solutions with VR exceeding 50\% were retained. Table~\ref{tab:mt_solution} summarizes the source parameters, nodal-plane geometry, moment tensor decomposition, and inversion quality metrics for the 37 accepted events, while Figure~\ref{fig:focal_mt} shows the spatial distribution of the corresponding deviatoric moment tensor solutions for the 13 events.

The distributions of strike, dip, and rake angles are summarized in Figure~\ref{fig:focal_mt}(b-d). The strike orientations define two dominant north--south trending populations that closely match the relocated seismicity clusters and support the interpretation of two subparallel fault strands. Dip angles cluster around $\sim30^\circ$ and $\sim60^\circ$, consistent with conjugate normal faults, whereas rake angles are concentrated near $270^\circ$, indicating predominantly normal faulting with only minor strike-slip components. 

Representative waveform fits are shown in SI Figure~\ref{fig:mtinv_example}. The examples correspond to the accepted solutions with the highest (65\%) and lowest (50\%) variance reductions. Synthetic waveforms closely reproduce the observed radial, transverse, and vertical components across the adopted frequency band, indicating that the inferred source mechanisms are consistent across the range of accepted inversion qualities.


\begin{table}[h!]
\centering
\caption{
Moment tensor solutions for earthquakes with $M_l > 3.0$ and variance reduction exceeding 50\%. The table lists the event origin time (Event ID), hypocentral location, magnitude, and the strike, dip, and rake parameters of the two nodal planes obtained from the moment tensor inversion. The percentage contributions of the double-couple (DC) and compensated linear vector dipole (CLVD) components of the deviatoric moment tensor solution are also reported.
}
\resizebox{\textwidth}{!}{%
\begin{tabular}{lcccccccc}
\hline
Event ID & Lat ($^\circ$N) & Lon ($^\circ$E) & Depth (km) & Fault Plane 1 (S/D/R) & Fault Plane 2 (S/D/R) & DC/CLVD/ISO (\%) & VR (\%) & $M_w$ \\
\hline
2019\_03\_01\_05\_44\_25 & 20.037 & 72.924 & 4.83 & 345/36/$-$103 & 181/55/$-$81 & 97/3/0 & 58.2 & 3.79 \\
2019\_03\_10\_08\_25\_26 & 20.054 & 72.908 & 1.64 & 1/57/$-$96 & 192/33/$-$81 & 98/2/0 & 52.9 & 2.74 \\
2019\_03\_10\_19\_30\_10 & 20.051 & 72.920 & 5.12 & 206/65/$-$67 & 341/34/$-$130 & 85/15/0 & 55.7 & 3.17 \\
2019\_03\_10\_22\_21\_49 & 20.055 & 72.908 & 1.56 & 335/58/$-$119 & 201/42/$-$52 & 68/32/0 & 51.3 & 2.76 \\
2019\_03\_31\_12\_21\_57 & 20.049 & 72.923 & 5.58 & 40/72/$-$73 & 176/25/$-$132 & 95/5/0 & 57.1 & 3.09 \\
2019\_05\_13\_06\_31\_26 & 20.046 & 72.921 & 6.32 & 180/31/$-$115 & 28/62/$-$76 & 84/16/0 & 59.0 & 2.99 \\
2019\_07\_24\_19\_48\_28 & 20.056 & 72.909 & 1.51 & 181/35/$-$90 & 1/55/$-$90 & 85/15/0 & 67.8 & 2.75 \\
2019\_07\_31\_14\_50\_03 & 20.085 & 72.907 & 1.18 & 113/37/$-$167 & 13/82/$-$54 & 93/7/0 & 59.2 & 2.86 \\
2019\_10\_24\_19\_12\_15 & 20.063 & 72.950 & 1.85 & 5/54/$-$84 & 175/37/$-$98 & 83/17/0 & 59.5 & 2.79 \\
2019\_10\_25\_22\_35\_41 & 20.020 & 72.908 & 4.51 & 201/60/$-$71 & 346/36/$-$120 & 50/50/0 & 59.8 & 3.27 \\
2019\_11\_19\_20\_05\_44 & 20.057 & 72.950 & 1.02 & 342/61/$-$99 & 180/30/$-$74 & 89/11/0 & 55.1 & 3.27 \\
2019\_11\_19\_20\_26\_47 & 20.059 & 72.949 & 0.82 & 173/32/$-$89 & 351/58/$-$91 & 94/6/0 & 55.1 & 2.79 \\
2019\_11\_19\_20\_57\_41 & 20.052 & 72.952 & 0.78 & 351/64/$-$95 & 183/26/$-$79 & 80/20/0 & 50.2 & 2.66 \\
2019\_11\_21\_01\_50\_28 & 20.070 & 72.953 & 2.60 & 193/43/$-$73 & 350/49/$-$106 & 100/0/0 & 50.2 & 3.19 \\
2019\_11\_23\_15\_10\_48 & 20.067 & 72.952 & 2.20 & 189/37/$-$78 & 354/54/$-$99 & 98/2/0 & 60.9 & 3.25 \\
2019\_11\_23\_20\_31\_13 & 20.067 & 72.951 & 1.99 & 344/53/$-$102 & 184/39/$-$74 & 91/9/0 & 60.2 & 3.26 \\
2019\_11\_24\_12\_12\_25 & 20.044 & 72.940 & 1.58 & 178/27/$-$82 & 349/63/$-$94 & 96/4/0 & 51.1 & 2.70 \\
2019\_11\_25\_21\_15\_55 & 20.039 & 72.945 & 1.58 & 178/27/$-$82 & 349/63/$-$94 & 96/4/0 & 51.1 & 2.70 \\
2019\_11\_27\_01\_25\_37 & 20.047 & 72.952 & 1.52 & 2/58/$-$95 & 190/32/$-$83 & 89/11/0 & 61.0 & 2.71 \\
2019\_11\_30\_11\_41\_33 & 20.081 & 72.941 & 1.42 & 358/62/$-$64 & 131/37/$-$130 & 62/38/0 & 54.6 & 3.01 \\
2019\_12\_03\_06\_09\_23 & 20.093 & 72.944 & 1.57 & 5/59/$-$78 & 164/33/$-$108 & 92/8/0 & 55.4 & 2.65 \\
2019\_12\_03\_22\_17\_43 & 19.964 & 72.889 & 1.87 & 13/58/$-$93 & 199/33/$-$85 & 87/13/0 & 50.1 & 2.66 \\
2019\_12\_13\_23\_52\_38 & 20.057 & 72.952 & 1.70 & 341/53/$-$105 & 185/40/$-$71 & 95/5/0 & 56.9 & 3.68 \\
2019\_12\_14\_09\_05\_05 & 20.101 & 72.950 & 1.24 & 9/59/$-$72 & 157/35/$-$117 & 97/3/0 & 52.1 & 2.70 \\
2019\_12\_14\_10\_47\_34 & 20.063 & 72.951 & 1.55 & 179/38/$-$94 & 4/52/$-$87 & 79/21/0 & 63.1 & 2.77 \\
2019\_12\_15\_00\_31\_28 & 20.015 & 72.920 & 2.91 & 202/65/$-$78 & 356/27/$-$114 & 94/6/0 & 55.7 & 3.03 \\
2019\_12\_15\_03\_49\_17 & 20.002 & 72.908 & 3.02 & 343/42/$-$144 & 225/67/$-$54 & 88/12/0 & 61.4 & 2.79 \\
2019\_12\_15\_19\_27\_56 & 20.049 & 72.922 & 4.63 & 181/89/98 & 277/8/6 & 96/4/0 & 59.2 & 2.82 \\
2020\_01\_31\_07\_54\_42 & 20.127 & 72.952 & 3.93 & 217/80/$-$61 & 324/30/$-$160 & 76/24/0 & 53.0 & 3.30 \\
2020\_02\_23\_04\_43\_02 & 20.067 & 72.952 & 1.92 & 177/43/$-$89 & 355/47/$-$91 & 95/5/0 & 53.2 & 2.73 \\
2020\_03\_19\_03\_49\_08 & 20.048 & 72.923 & 5.87 & 15/68/$-$66 & 144/32/$-$136 & 91/9/0 & 52.1 & 3.09 \\
2020\_04\_01\_18\_09\_18 & 20.044 & 72.921 & 6.29 & 165/31/$-$125 & 25/65/$-$71 & 92/8/0 & 55.5 & 3.01 \\
2020\_04\_05\_18\_48\_30 & 20.131 & 72.947 & 2.63 & 210/44/$-$46 & 337/60/$-$124 & 79/21/0 & 50.3 & 3.21 \\
2020\_09\_10\_22\_27\_48 & 20.040 & 72.909 & 3.06 & 11/58/$-$93 & 196/32/$-$85 & 91/9/0 & 53.6 & 3.39 \\
2020\_09\_11\_01\_36\_02 & 20.033 & 72.907 & 1.92 & 7/57/$-$88 & 184/33/$-$93 & 80/20/0 & 65.7 & 3.23 \\
2020\_09\_21\_21\_20\_44 & 20.050 & 72.922 & 2.87 & 201/34/$-$72 & 359/57/$-$102 & 86/14/0 & 58.5 & 2.90 \\
2020\_09\_22\_19\_11\_15 & 20.038 & 72.904 & 1.79 & 19/62/$-$90 & 200/28/$-$89 & 64/36/0 & 51.0 & 2.78 \\
\hline
\end{tabular}
}\label{tab:mt_solution}
\end{table}


\begin{figure}[h!]
    \centering
    \includegraphics[width=0.8\linewidth]{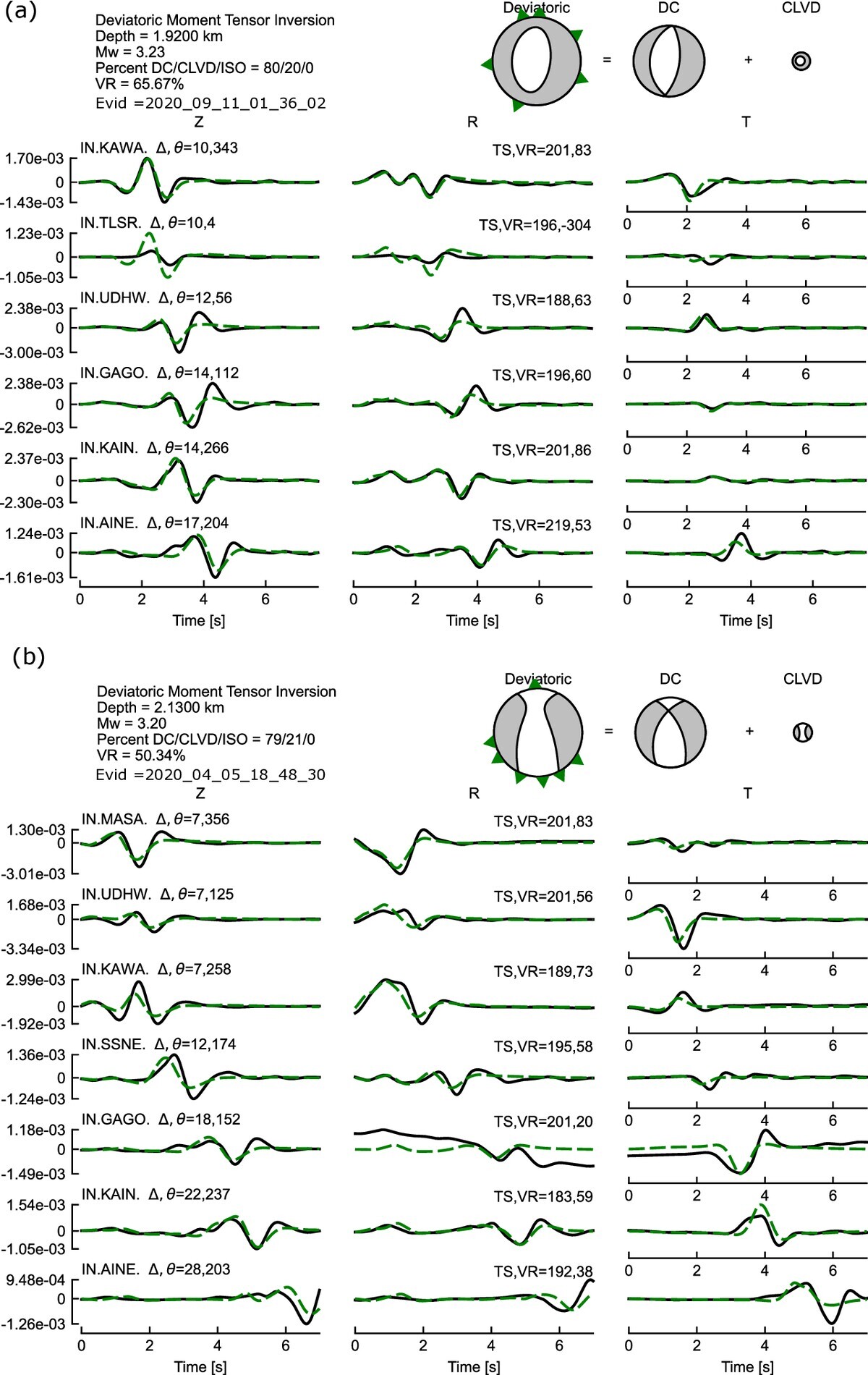}
    \caption{Deviatoric moment tensor inversion results for two Palghar earthquakes filtered between 0.05--1 Hz. 
    (a) Event on 2020-09-11 ($M_w$ 3.23) with the highest variance reduction (VR = 65\%). 
    (b) Event on 2020-04-05 ($M_w$ 3.21) with the lowest variance reduction (VR = 50\%). 
    Black lines are observed waveforms, green lines are synthetics, shown for radial (R), transverse (T), and vertical (Z) components. 
    Beachball diagrams show deviatoric solutions, with percentage decomposition and overall VR listed above each event. 
    }
    \label{fig:mtinv_example}
\end{figure}
\section*{Text S4.  Influence of Catalog Resolution on Inferred Swarm Migration}
The migration behavior inferred from the Palghar swarm depends strongly on the accuracy and density of the earthquake catalog. SI Figure~ \ref{fig:diffusion_same_color} compares the distance--time evolution obtained from the initial human-picked catalog containing 10,114 events with the migration envelopes derived from the relocated catalog used in this study. Although the initial catalog captures the broad temporal expansion of the swarm, the substantial scatter in earthquake locations obscures the detailed migration fronts and makes it difficult to distinguish the different phases of swarm evolution. In particular, the delayed activation of the eastern fault, the rapid migration episodes, and the temporal evolution of the leading seismicity fronts are not clearly resolved.

Application of \texttt{NonLinLoc} to the densified ML-enhanced catalog improves the overall spatial coherence of the seismicity and captures the first-order expansion of the swarm (SI Figure~\ref{fig:diffusion_nlloc}). The eastern, western, and off-fault seismicity clusters become more readily distinguishable, and the broad diffusive character of the swarm is preserved. However, the location precision remains insufficient to clearly resolve the fault-scale migration fronts, persistent seismicity bands, and episodic along-strike and along-dip migration patterns revealed by the subsequent \texttt{GrowClust} relocations.

The \texttt{GrowClust} relocations further sharpen the seismicity distribution and reveal the detailed spatiotemporal patterns analyzed in the main text. The improved relative locations delineate persistent localized seismicity along the western fault, distinct migration episodes along both fault systems, and the close correspondence between on-fault and off-fault migration patterns. These comparisons demonstrate that the combination of catalog densification and high-precision relative relocation is essential for understanding the processes governing the evolution of the Palghar swarm.
\begin{figure}[ht]
    \centering
    \includegraphics[width=1.0\linewidth]{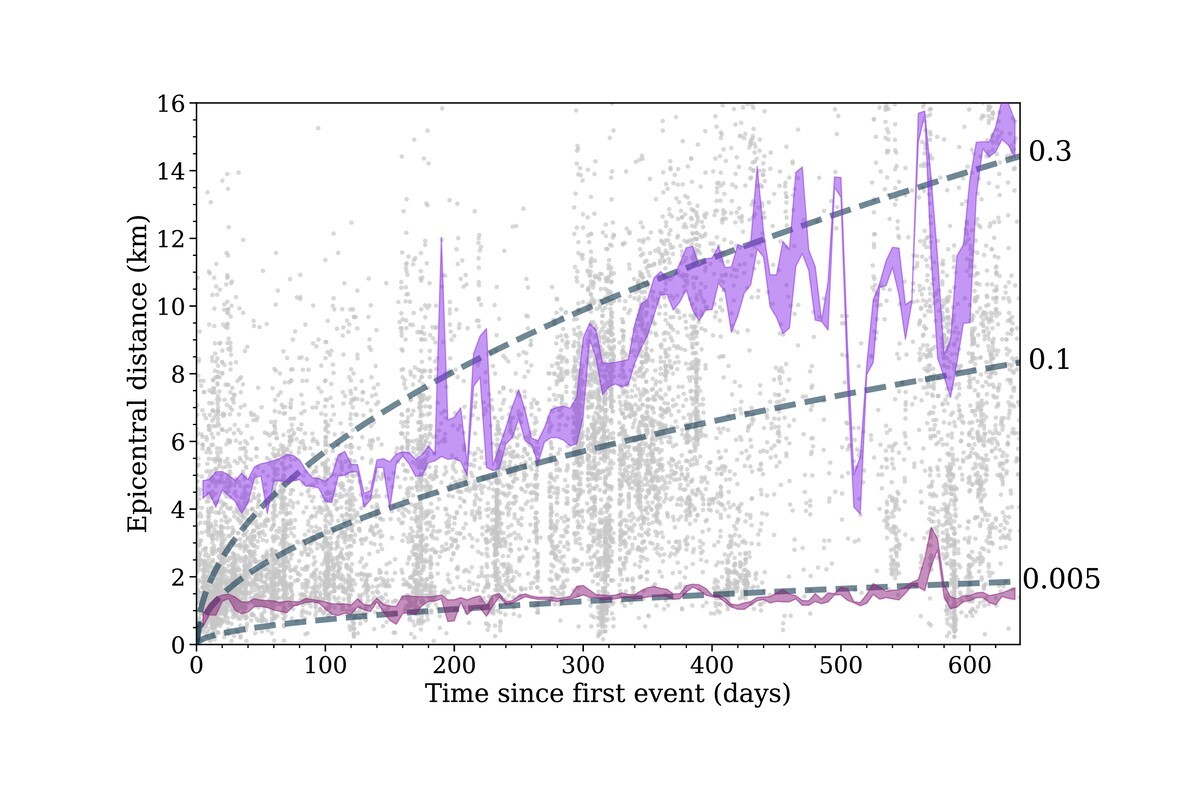}
    \caption{
    Epicentral distance as a function of time since the first detected event for the 10,114 earthquakes in the initial human-picked catalog (gray dots). Dashed curves indicate theoretical diffusion fronts corresponding to apparent diffusivities of $D = 0.005$, 0.1, and 0.3~m$^{2}$~s$^{-1}$. The purple and pink shaded regions show the 95th and 5th percentile migration envelopes obtained from the relocated catalog, with shaded bounds denoting the associated $\pm$2.5\% percentile range. The figure illustrates that the diffuse nature of the initial earthquake locations obscures the detailed migration fronts and spatiotemporal patterns that become evident only after relocation.}
    \label{fig:diffusion_same_color}
\end{figure}

\begin{figure}[ht]
    \centering
    \includegraphics[width=1.0\linewidth]{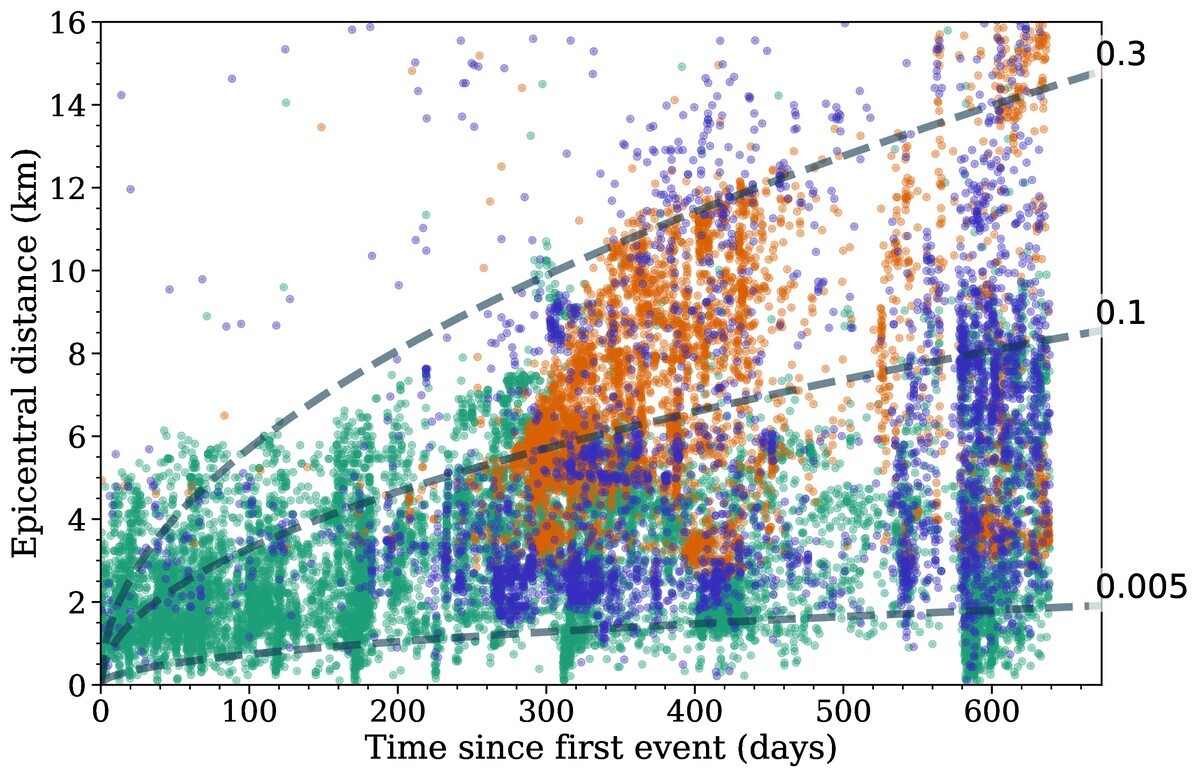}
    \caption{Distance--time evolution of the Palghar swarm using 20,638 earthquakes located with \texttt{NonLinLoc} and recorded by six or more stations. Green, orange, and blue dots represent earthquakes associated with the western cluster, eastern cluster, and off-fault seismicity, respectively.  Epicentral distance from the first detected event is plotted as a function of time since swarm initiation. Dashed curves represent reference diffusion fronts corresponding to apparent diffusivities of 0.005, 0.1, and 0.3~m$^{2}$~s$^{-1}$. The figure illustrates the broad spatiotemporal expansion of the swarm resolved by the \texttt{NonLinLoc} locations and serves as a reference for comparison with the higher-resolution catalog obtained from \texttt{GrowClust} relocations using waveform cross-correlation coefficients $\geq 0.8$.}
    \label{fig:diffusion_nlloc}
\end{figure}

\begin{figure}[ht]
    \centering
    \includegraphics[width=0.8\linewidth]{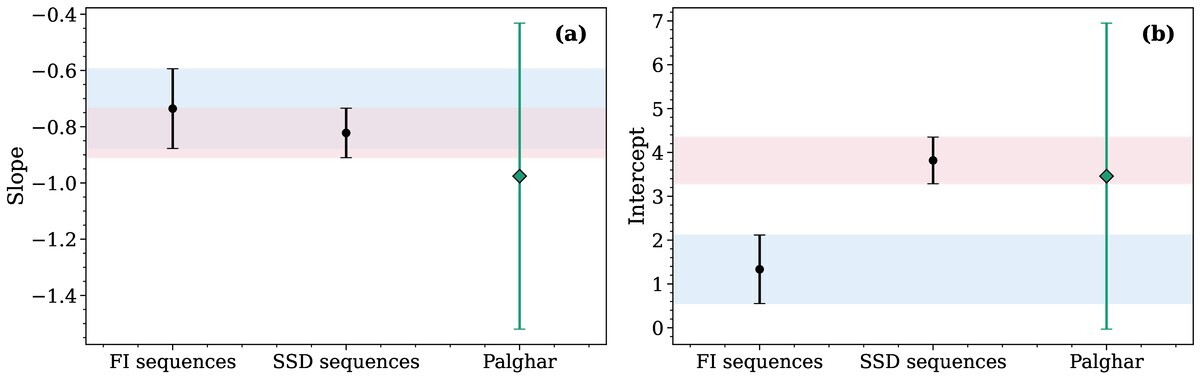}
    \caption{Comparison of velocity--duration scaling parameters for the Palghar swarm with fluid-induced (FI) and slow-slip-dominated (SSD) sequences compiled by \citeA{danre2024parallel}. (a) Slope and (b) intercept for the Palghar swarm (green diamonds) and reference sequences (blue and red, respectively). The Palghar parameters are derived from migration episodes within the hatched region in Figure~\ref{fig:migration}e. Error bars denote 95\% confidence intervals calculated assuming Gaussian errors.}
    \label{fig:danre_stats}
\end{figure}

\section*{Text S5. Along-Strike, Along-Dip, and Off-Fault Migration Patterns of the High-Resolution Relocated Seismicity}\label{tex:earthquake_migration}
To further investigate the migration behavior of the Palghar swarm, we analyzed the temporal evolution of seismicity projected along the strike and dip directions of the western and eastern faults. Earthquake locations were projected onto fault-parallel coordinate systems, and the evolution of the migrating seismicity fronts was tracked using the 95th and 5th percentile envelopes of seismicity through time. Migration velocities were estimated by linear regression of the corresponding percentile fronts over the selected time intervals, with the numbered episodes in Figures~\ref{fig:along_trace_distance_west} and \ref{fig:along_trace_distance_east} marking the individual migration phases analyzed. The hatched region in Figure~\ref{fig:migration}e identifies the group of faster migration episodes selected for the velocity-duration scaling analysis and subsequent slope-intercept comparison \ref{fig:danre_stats}.

For the western fault, seismicity remains strongly concentrated within persistent central bands located near $\sim$1.2 km in the along-dip direction and $\sim$1.6 km in the along-strike direction. To better resolve the migration of the seismicity fronts, events within these highly populated bands were excluded from the percentile calculations using windows of $\pm$0.5 km and $\pm$0.7 km in the along-dip and along-strike projections, respectively. The resulting distance-time evolution reveals strongly episodic migration, characterized by multiple propagating fronts, abrupt reversals, and back-propagating seismicity streaks along both strike and dip directions. A prominent down-dip migration episode is observed between approximately 220 and 450 days (migration 7 in Figure~\ref{fig:along_trace_distance_west}a), followed by a later phase of up-dip and southward migration between approximately 480 and 640 days (migrations 9 and 15 in Figure~\ref{fig:along_trace_distance_west}a,c). Several additional shorter-duration migration episodes occur throughout the sequence, indicating repeated reorganization of seismicity along the fault. 

In contrast, the eastern fault exhibits a more spatially localized migration pattern. Following its activation at approximately 260 days, seismicity remains concentrated near the central portion of the fault before progressively expanding along the dip direction. After approximately 400 days, the seismicity envelope broadens, with expansion toward updip and a pronounced down-dip migration of the lower migration front during the later stages of the sequence (migration 2 in Figure~\ref{fig:along_trace_distance_east}a). In the along-strike projection, the leading migration front propagates northward between approximately 260 and 460 days (migrations 3 and 4 in  Figure~\ref{fig:along_trace_distance_east}c). At the same time, a weaker migration front develops toward negative along-strike distances (migration 7).

To further examine the relationship between on-fault and off-fault seismicity, we compared the epicentral migration of the off-fault events with that of the eastern fault (SI Figure~\ref{fig:off_fault_500_m} a, d). Between approximately 260 and 460 days, both clusters exhibit nearly identical migration rates. During the initial stage of this interval (approximately 260--360 days), the off-fault seismicity also reproduces the northward migration pattern observed along the eastern fault (SI Figure~\ref{fig:off_fault_500_m} b, c). This correspondence becomes even more apparent when the on-fault clusters are restricted to narrower bands of $\pm$500~m around the interpreted fault traces (SI Figure~\ref{fig:off_fault_500_m} e, f), demonstrating that the observed migration patterns are robust and insensitive to the choice of fault-zone width.  During the later stages of the sequence (approximately 480--640 days), the southward and up-dip migration observed along the western fault is similarly mirrored by adjacent off-fault seismicity immediately west of the fault (SI Figure~\ref{fig:off_fault_migration}). These observations demonstrate that the off-fault seismicity closely tracks the migration of the active fault systems throughout the evolution of the swarm.

\begin{figure}[h!]
    \centering
    \includegraphics[width=1.0\linewidth]{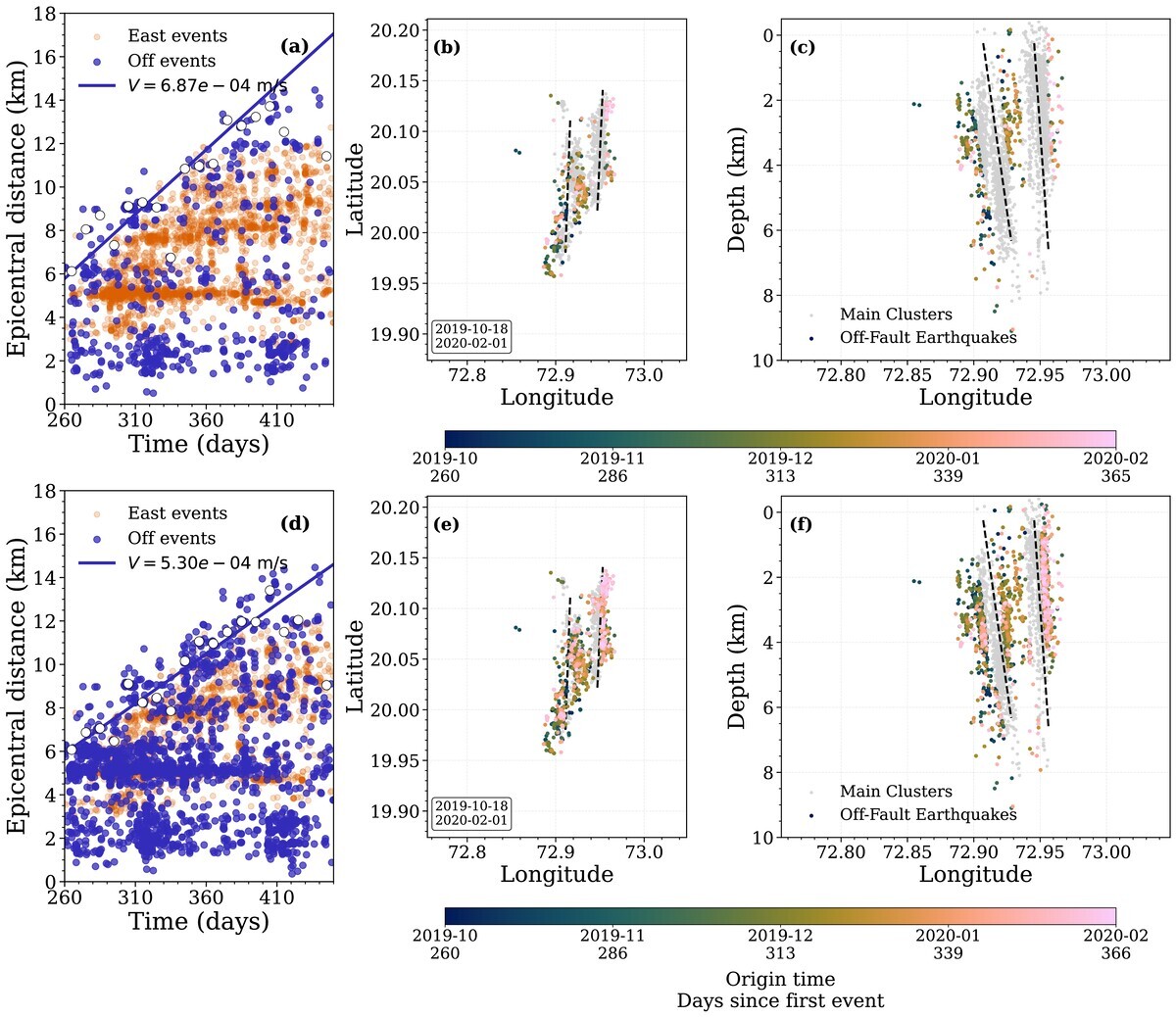}
    \caption{
    Migration of off-fault seismicity associated with the rapid migration episode on the eastern fault (migration 3 in Figure~\ref{fig:along_trace_distance_east}). (a) Epicentral distance as a function of time for off-fault earthquakes identified using a $\pm$1 km fault-zone definition. Orange and blue dots represent eastern-fault and off-fault earthquakes, respectively. The solid blue line shows a linear fit to the leading migration front of the off-fault seismicity, corresponding to a migration velocity of $6.87 \times 10^{-4}$~m~s$^{-1}$. (b) Map-view and (c) longitude--depth projections of the seismicity during the 260--360 day interval. Gray dots represent on-fault earthquakes associated with the eastern and western fault systems, while colored dots denote off-fault earthquakes color-coded by origin time. The off-fault seismicity exhibits the same northward migration pattern observed along the eastern fault (migration 3 in Figure ~\ref{fig:along_trace_distance_east}). (d--f) Same as (a--c), but using a stricter on-fault definition in which earthquakes within $\pm$500 m of the interpreted eastern and western fault planes are classified as on-fault events, with all remaining earthquakes assigned to the off-fault population. The estimated off-fault migration velocity ($5.30 $$\times 10^{-4}$~m~s$^{-1}$) and the spatiotemporal migration pattern remain largely unchanged, indicating that the observed off-fault migration is robust and not sensitive to the fault-zone width used to classify earthquakes.}
    \label{fig:off_fault_500_m}
\end{figure}

\begin{figure}[ht]
    \centering
    \includegraphics[width=1.0\linewidth]{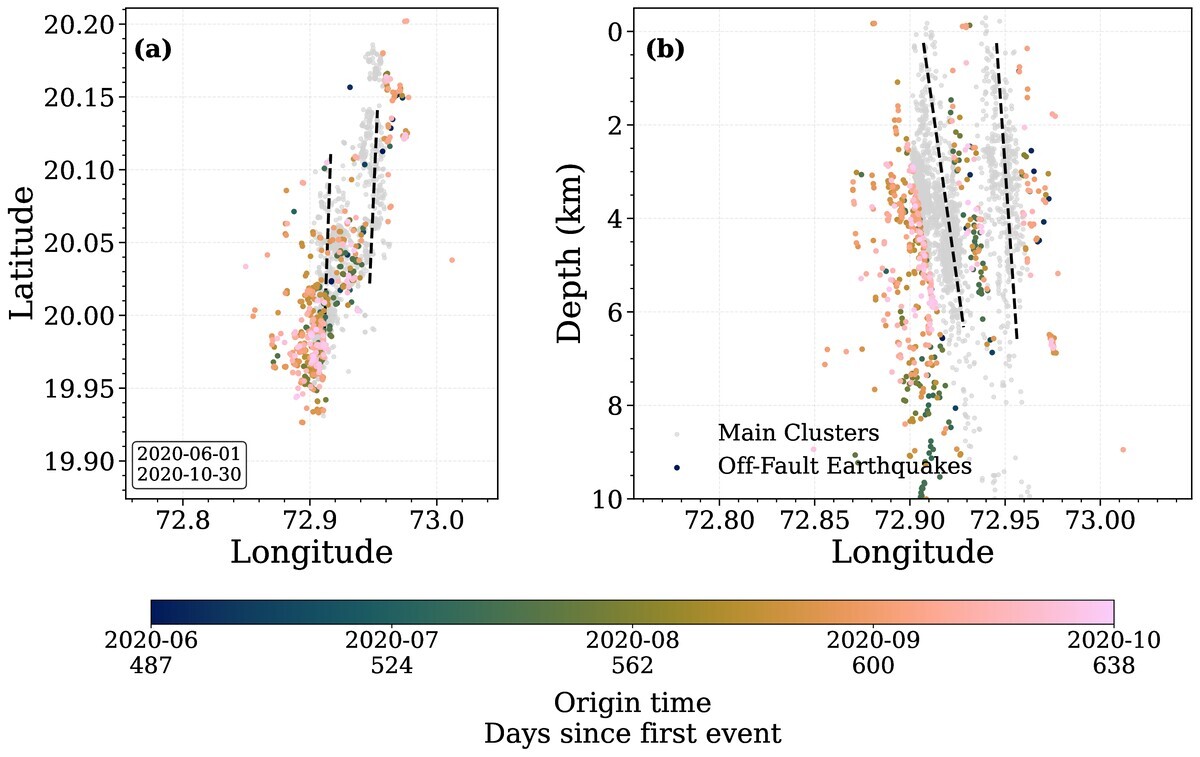}
    \caption{ 
    Spatiotemporal evolution of off-fault seismicity during the late stages of the Palghar swarm sequence (2020-06-01 to 2020-10-30, 480 to 640 days). Gray dots represent earthquakes associated with the western and eastern faults, while colored dots denote off-fault earthquakes, color-coded by origin time. (a) Map view and (b) longitude--depth projection of the off-fault seismicity. Dashed gray lines indicate the approximate geometry of the western and eastern fault planes inferred from relocated seismicity. The off-fault earthquakes exhibit migration patterns that closely mirror contemporaneous migration along the main fault systems, including the up-dip expansion observed on both the western and eastern faults and the southward migration episode identified on the western fault. These migrations correspond to episodes 9 and 15 on the western fault and episode 2 on the eastern fault (see Figures~\ref{fig:along_trace_distance_west} and \ref{fig:along_trace_distance_east}), highlighting the close spatiotemporal relationship between on-fault and off-fault seismicity.}
    \label{fig:off_fault_migration}
\end{figure}

\begin{figure}[ht]
    \centering
    \includegraphics[width=1.0\linewidth]{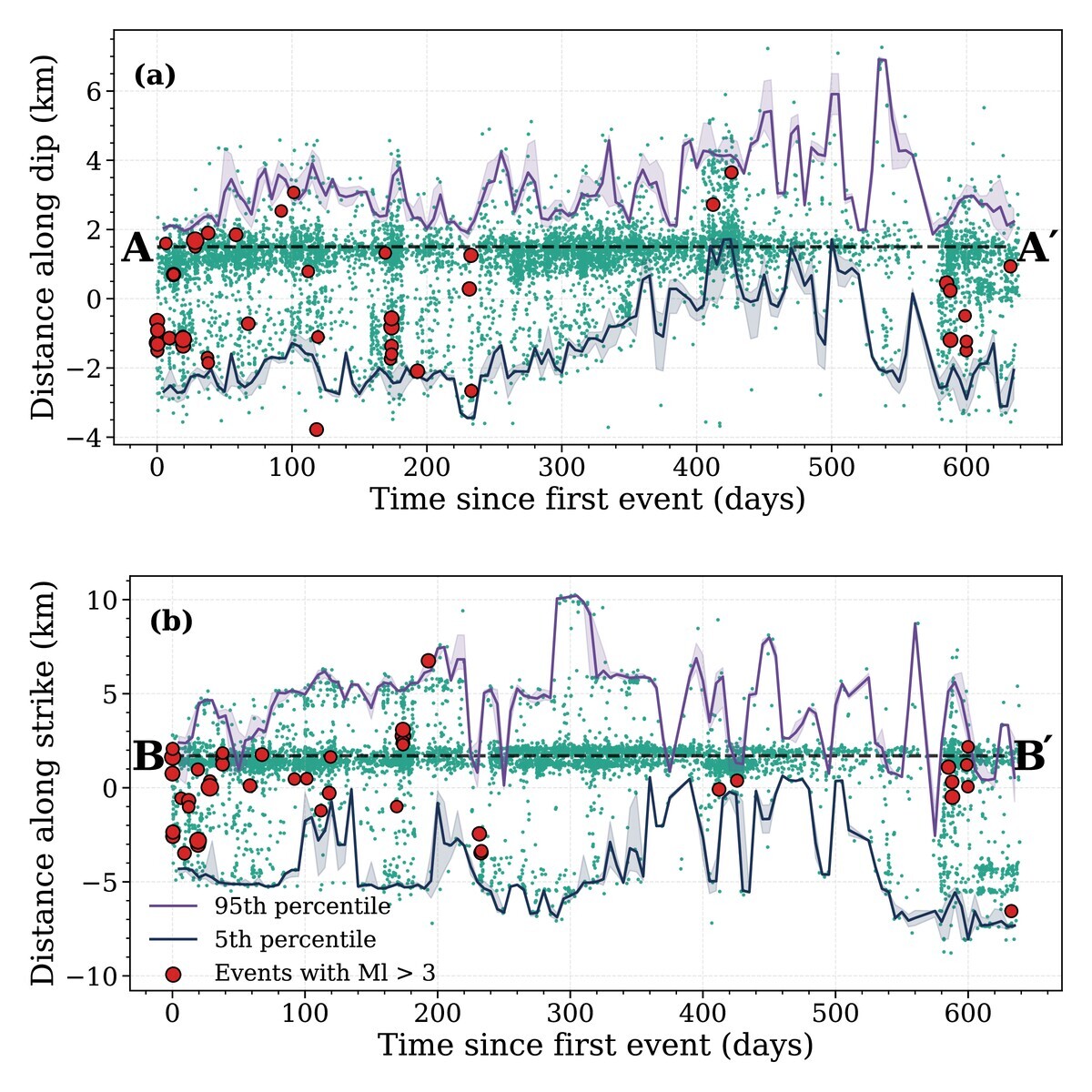}
    \caption{
    Spatiotemporal evolution of seismicity along the western fault after restricting the fault-zone width to $\pm$500 m around the interpreted fault plane. Earthquakes are projected onto the along-dip (a) and along-strike (b) directions as a function of time since the first detected event. Green dots represent relocated earthquakes, while red circles denote larger events with local magnitude ($M_l$$>$  3). The horizontal dashed lines mark the dominant seismicity band. Purple and dark-blue curves show the 95th and 5th percentile positions of seismicity, respectively, with shaded regions indicating the associated $\pm$2.5\% percentile bounds. Despite the substantially narrower fault-zone definition, a persistent and spatially confined seismicity band remains evident in both the along-dip and along-strike projections throughout most of the swarm evolution, demonstrating that the long-lived seismicity concentration along the western fault is a robust feature and not an artifact of the fault-zone width used for event classification.}
    \label{fig:western_fault_500_m}
\end{figure}


\begin{figure}[ht]
    \centering
    \includegraphics[width=1.0\linewidth]{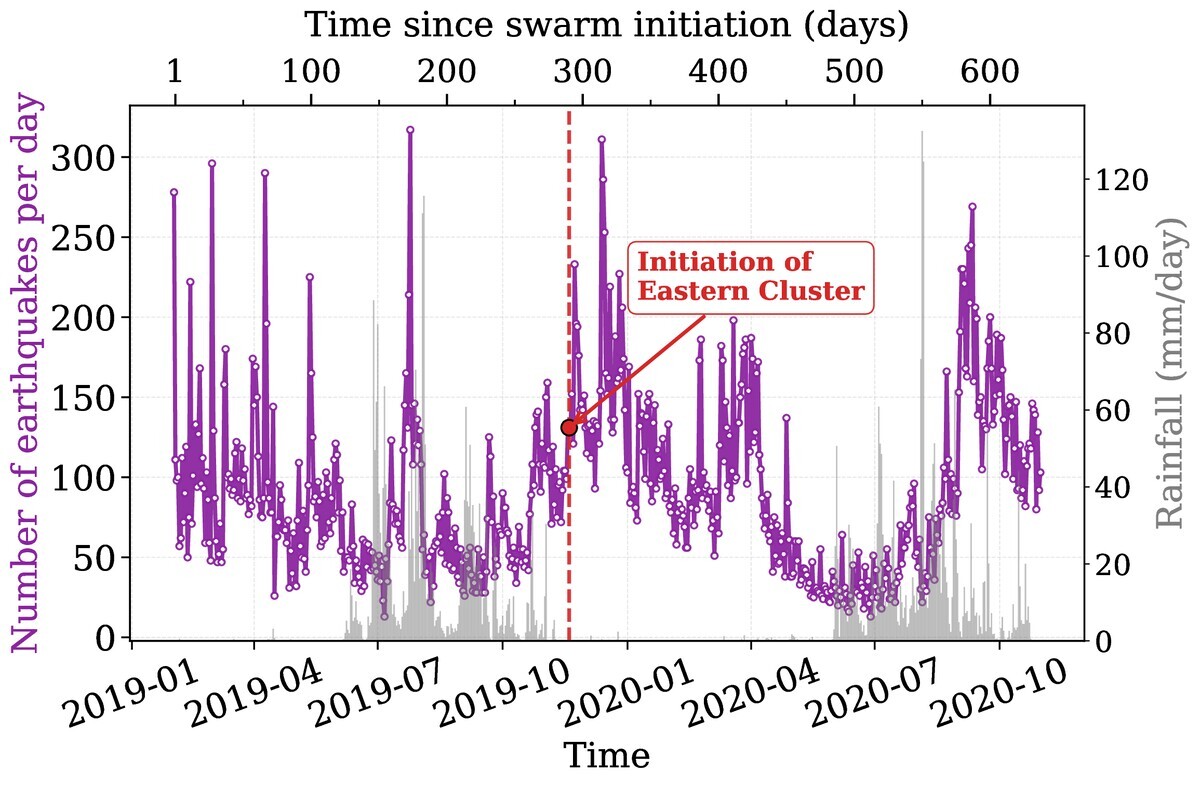}
    \caption{
    Time evolution of daily earthquake frequency during the Palghar swarm sequence from February 2019 to October 2020, comprising a total of 57,826 earthquakes detected in the ML-based earthquake catalog developed in this study. The purple curve shows the number of earthquakes per day, while gray bars represent daily rainfall. The red dashed line marks the onset of seismicity on the eastern fault. The periods of highest seismicity rates occur during the early stages of the sequence and do not show an obvious correspondence with seasonal rainfall variations.  Daily precipitation data were obtained from the NASA Langley Research Center Prediction Of Worldwide Energy Resources (POWER) Project.}
    \label{fig:events_time_series}
\end{figure}



\clearpage
\renewcommand*{\bibfont}{\normalsize}
\printbibliography
\end{document}